\documentclass[10pt,twocolumn]{article}

\usepackage[a4paper,
left=1.55cm,
right=1.55cm,
top=2cm,
bottom=2cm]{geometry}

\usepackage{amsmath,amssymb,amsfonts}
\usepackage{mathtools}
\usepackage{microtype}

\usepackage{algorithmic}
\usepackage{algorithm}
\usepackage{array}
\usepackage[caption=false,font=normalsize,labelfont=sf,textfont=sf]{subfig}
\usepackage{xcolor}
\usepackage{textcomp}
\usepackage{stfloats}
\usepackage{url}
\usepackage{verbatim}
\usepackage{graphicx}
\usepackage{cite}
\usepackage{tikz}
\usepackage{abstract}
\usepackage{titlesec}
\usepackage{fancyhdr}
\begin{document}
 
\twocolumn[{%
\begin{@twocolumnfalse}
 
\title{\textbf{Taking advantage of multiple scattering for Optical Reflection Tomography}}
 
\author{Thomas Wasik$^{1,2}$, Victor Barolle$^{3}$, Alexandre Aubry$^{1}$ and Josselin Garnier$^{2}$\\[6pt]
\small $^{1}$Institut Langevin, ESPCI Paris, PSL University, CNRS, 75005 Paris, France\\
\small $^{2}$CMAP, Ecole polytechnique, Institut Polytechnique de Paris, 91120 Palaiseau, France\\
\small $^{3}$OWLO SAS, 75014 Paris, France\\[4pt]
\small \texttt{thomas.wasik@polytechnique.edu} \\
\small \texttt{victor@owlo.io} \\
\small \texttt{alexandre.aubry@espci.fr} \\
\small \texttt{josselin.garnier@polytechnique.edu}
}
 
\date{%
\small \copyright~2026 IEEE. Personal use of this material is permitted. Permission from IEEE must be obtained for all other uses, in any current or future media, including reprinting/republishing this material for advertising or promotional purposes, creating new collective works, for resale or redistribution to servers or lists, or reuse of any copyrighted component of this work in other works.\\[6pt]
\small \textit{Published in IEEE Transactions on Computational Imaging.}
}
 
\maketitle
 
\begin{abstract}
Optical Diffraction Tomography (ODT) is a powerful non-invasive imaging technique widely used in biological and medical applications. While significant progress has been made in transmission configuration, reflection ODT remains challenging due to the ill-posed nature of the inverse problem. We present a novel optimization algorithm for 3D refractive index (RI) reconstruction in reflection-mode microscopy.
Our method takes advantage of the multiply-scattered waves that are reflected by uncontrolled background structures and that illuminate the foreground RI from behind.
It tackles the ill-posed nature of the problem using weighted time loss, positivity constraints and Total Variation regularization.
We have validated our method with data generated by detailed 2D and 3D simulations, demonstrating its performance under weak multiple scattering conditions and with simplified forward models used in the optimization routine for computational efficiency. In addition, we highlight the need for multi-wavelength analysis and the use of regularization to ensure the reconstruction of the low spatial frequencies of the foreground RI.
\end{abstract}
 
\vspace{4pt}
\noindent\textbf{Keywords:} Reflection optical phase tomography, ill-posed inverse problem, multiple scattering, total variation regularization, temporal loss function, stochastic proximal-gradient.
\vspace{10pt}
 
\end{@twocolumnfalse}
}]

\section{Introduction}
Optical diffraction tomography (ODT) is a non-invasive quantitative imaging technique \cite{firstborn, choi2007tomographic, jin2017tomographic} that has shown promising applications in biology and medical imaging \cite{liu2016cell, park2018quantitative}. This technique aims to reconstruct the 3D refractive index (RI) map of a sample by illuminating it from various angles, recovering the complex scattered field using holography techniques \cite{kim2010principles}, and numerically solving the inverse scattering problem of light. This inverse problem is known as an ill-posed problem in transmission and in reflection mode configuration.
\\
ODT was first applied to weakly scattering systems, which can be described using the first Born and Rytov approximations \cite{firstborn, devaney1981inverse}. Under first Born hypothesis, the measurements present a filtering effect \cite{firstborn}.  In a transmission configuration, the optical system acts as a low-pass filter of the spatial frequencies of the object along the optical axis, corresponding to the missing-cone problem, while it acts as a high-pass filter in a reflection configuration \cite{coupland2008holography}. In recent years, ODT techniques have primarily been developed for transmission configurations, with significant efforts focused on simulating stronger scattering regimes and mitigating the missing-cone problem by developing new forward models and iterative techniques for solving the inverse problem \cite{ma2017optical, chen2020multi, lee2022inverse, pham2020three, kamilov2017plug}.
\\
The application of ODT methods to reflection microscopy would represent a significant breakthrough in the field of non-invasive imaging. Reflection microscopy, in particular, allows for the imaging of thicker samples and enables in vivo imaging \cite{weissleder2001clearer}.
ODT in reflection presents substantial challenges due to the ill-posed nature of the associated inverse problem. In the single scattering regime, only the high spatial frequencies of the object along the optical axis can be reconstructed \cite{coupland2008holography}. On the other hand, in the strongly scattering regime, the forward model becomes non-linear with respect to the RI map \cite{born2003principles} and the associated optimization problem is non-convex, providing no guarantee of convergence when employing gradient descent methods, which are commonly used in the literature \cite{ma2017optical, chen2020multi, lee2022inverse, pham2020three, kamilov2017plug}.
Several techniques have been developed for reflection ODT; however, they are limited by their conditions of application, either requiring a 1D reconstruction \cite{kang2022reflection} or relying on specific setups that do not allow for in vivo imaging \cite{li2024reflection}.
\\
Here, we propose a novel reconstruction algorithm to solve the 3D inverse problem, applicable to interferometric imaging in a reflection 
configuration with multi-wavelength incident wave-fields ~\cite{Lee2023,Zhang2023,balondrade2024multi}. The originality of our method lies in the necessity to account for multiple scattering phenomena in order to accurately reconstruct the RI map. It is intuitively based on the idea that, in the multiple scattering regime, the scattering medium itself becomes a source, making it possible to reconstruct a foreground object using the light reflected from the background. The solution to the inverse problem relies on an iterative optimization method that focuses on the low-frequency information of the foreground object. To guide the optimization, special time gating and regularization \cite{kamilovtv} on the loss will be applied. Our method shares similarities with techniques already explored in microscopy \cite{li2024reflection, kang2023tracing}, ultrasound~\cite{Weber2021,Stahli2021} and seismology~\cite{Mora1989,malcolm2009seismic}, 
where multiply-scattered waves enable the illumination of structures from behind and can give access to their low-spatial frequency RI distribution in a reflection configuration.
In contrast with the existing reflection-ODT method, our approach reconstructs the medium without relying on a known reflector, making it a promising candidate for in vivo imaging.
It differs from other reflection-based approaches, which reconstruct layers progressively, imaging deeper structures step by step \cite{layerstripping}.

\section{Multiple-scattering and filtering effect}

Reconstructing RI maps is an ill-posed problem: for similar measured fields, many different RI distributions can be valid solutions. 
In transmission, this leads to the missing cone problem, which elongates the reconstructed structures. In reflection, the measured signal corresponds mainly to the high spatial frequencies of the sample: the average RI may no longer be detectable. 

In this section, we show that, while this limitation holds under the first-order Born approximation, multiple scattering mitigates this effect (even at second order) and constrains the set of possible solutions of our inverse problem. In particular, we demonstrate that reflections from background structures provide additional information about foreground objects, enabling them to be reconstructed more accurately.

\subsection{Measurement as a filtering process}

In this paper, we will consider the following experimental configuration~\cite{Lee2023,Zhang2023,balondrade2024multi}: (\textit{i}) a biological medium is illuminated through a microscope objective (MO) by a set of incident wavefields at different wavelengths; (\textit{ii}) for each illumination, the backscattered wave field is collected through the same MO and interferes with a reference beam on a camera; (\textit{iii}) the amplitude and phase of each scattered wavefield is then extracted by an interferometric technique (phase shifting interferometry, on-axis/off-xis holography \textit{etc.}). Our goal is to retrieve from the set of recorded wave-fields the RI distribution inside the biological medium.

A biological medium can be described by an inhomogeneous scattering potential $V(x, y, z) = k_0^2 (n^2(x,y,z)-n_\text{b}^2)$, where $k_0$ is the wavenumber in vacuum, $n_\text{b}$ is the background RI and $n(x,y,z)$ is the RI map that we want to recover.  
The total field $E$ resulting from interaction with the RI map $n(x,y,z)$ satisfies Helmholtz equation:
\begin{align}
\label{Helmholtz}
(\nabla^2 + k_0^2 n^2(x,y,z))E(x,y,z) = 0
\end{align}

Let us express the complex wavefield as $E = E_{\text{in}}+E_\text{s}$ \cite{firstborn}, a sum of incident wave solution of the homogeneous Helmholtz equation $E_{\text{in}}$ and a scattered field $E_\text{s}$ solving:
\begin{align}
\label{Helmholtz scattered}
&(\nabla^2 + k_0^2 n_\text{b}^2)E_\text{s}(x,y,z) = U(x, y, z) 
\end{align}
with Sommerfeld radiation condition.
The equation \eqref{Helmholtz scattered} is the homogeneous Helmholtz equation with an internal source term $U(x,y,z)=V(x,y,z)E(x,y,z)$ \cite{coupland2008holography} and it can be solved with the Green's theorem which links the scattered field with the internal source thanks to the homogeneous three-dimensional Green's function $G(x, y, z) = {-e^{jn_\text{b}k_0\sqrt{x^2+y^2+z^2}}}/{(4 \pi \sqrt{x^2+y^2+z^2})}$:
\begin{align}
\label{convgreen}
E_\text{s}(&x,y,z) =  \iiint_{\mathbb{R}^3} U(x', y', z') \nonumber \notag \\
&\quad \times G(x-x', y-y', z-z') \, \text{d}x'\text{d}y'\text{d}z'
\end{align}

\begin{figure}[!t]
    \centering
    \includegraphics[width=1.0\linewidth]{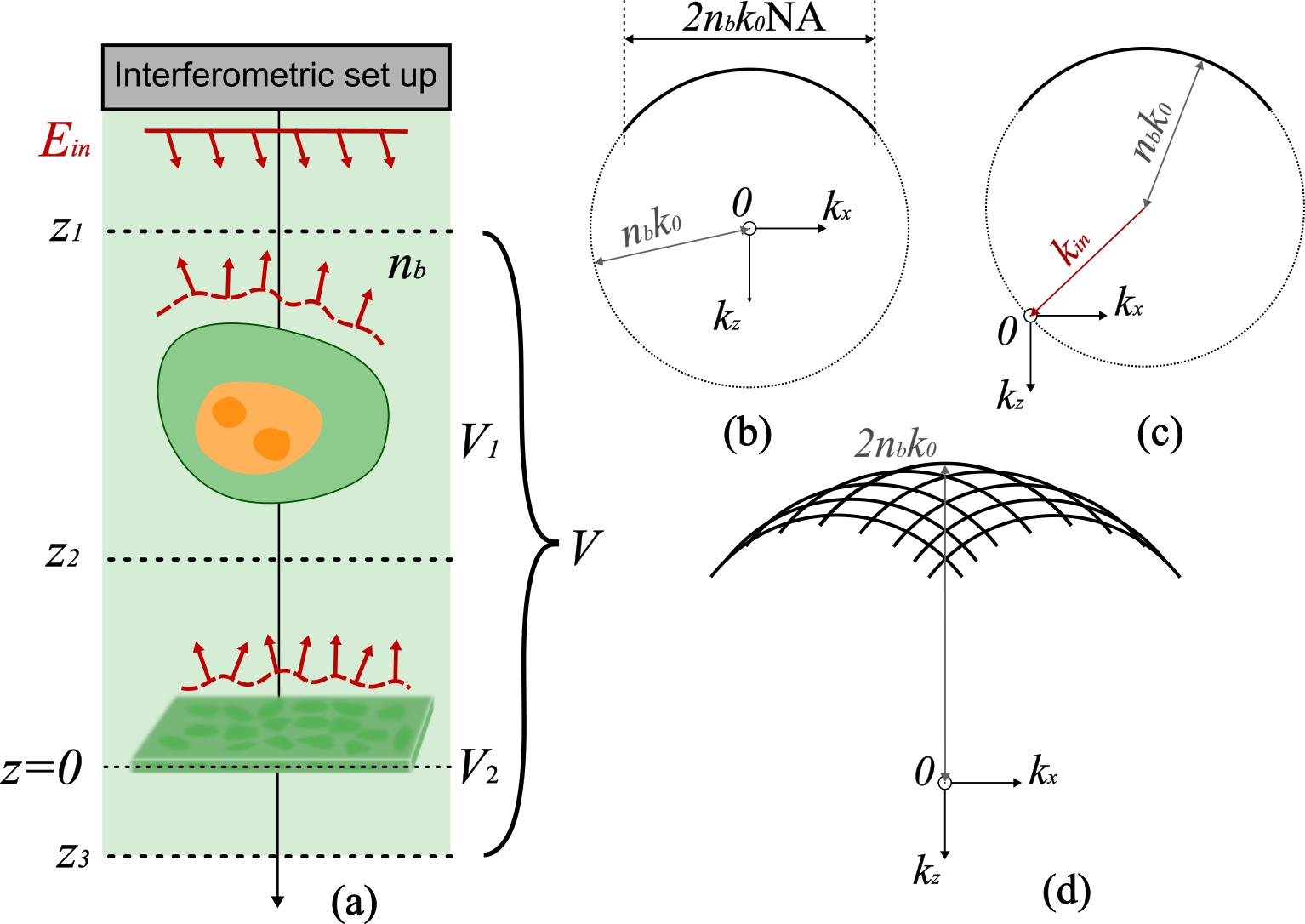}
    
    \caption{ Schematic representation of the medium to be imaged and illustration of the filtering effects.
(a) Illustration of the reflection configuration: the measurement system is positioned at 
$z<0$, while the incident waves propagate along the positive 
$z$-direction. The medium is separated into 2 parts: an object to be imaged with potential $V_1$ and a reflector with potential $V_2$.
(b) Description of the filtering operation in the (spatial) Fourier domain: only frequencies of $U=VE$ within the semicircle contribute to the measurement. The limited numerical aperture (NA) of the system restricts the acquisition to a portion of the Ewald half-sphere.  
(c) Filtering effect on $V$ in reflection under the first order Born approximation: the filter is located at high $z$-frequencies domain in a reflection configuration. 
(d) Effect of multiple illuminations on $V$ reconstruction under first order Born hypothesis: by multiplying the number of incident waves, filter functions are superposed, providing more information on the object. However, low spatial frequencies remain inaccessible.}
    \label{fig_1}
\end{figure}

In order to express the filtering process operated by the optical system, it is common to apply the first order Born approximation directly to $U$ ($U=VE_{\text{in}}$) in order to obtain a linear relation between the scattered wavefield $E_\text{s}$ and the scattering potential $V$. The filtering effect of the numerical aperture (NA) then directly applies to $V$.
However, in the multiple scattering regime, we emphasize that this filtering operation shall be applied to $U$ and not directly to $V$.
\\
As a convention and as illustrated in Fig.~\ref{fig_1}(a), the optical axis is here the $z$-axis and the measurement of the scattered wavefield is carried out at negative $z$.
If evanescent waves are neglected, the expression of the scattered field \eqref{convgreen} can be simplified (Appendix \ref{appendix:A}): 

\begin{align}
\label{Ewald sphere}
\mathcal{F}_{2D}[E_\text{s}(\cdot,z)](k_{x}, k_{y}) \propto \tilde{U}\left (k_x, k_y, \mathrm{sgn}(z)\sqrt{n_\text{b}^2k_0^2-k_x^2-k_y^2} \right )
\end{align}
where $\mathcal{F}_{2D}$ corresponds to the 2D Fourier transform and the symbol $\tilde{\cdot}$ stands for the 3D Fourier transform. Equation \eqref{Ewald sphere} means that it is only possible to recover information in the Ewald sphere of the source $U$ in Fourier space (Fig.~\ref{fig_1}(b)). In our case, the measurement system is placed in negative $z$: only the lower half sphere can be retrieved.

Experimentally, it is not possible to measure the entire spatial frequency spectrum of $E_{\text{s}}(\cdot,z)$, meaning its high-frequency components remain inaccessible in practice. This results in a filtering effect on $E_{\text{s}}(.,z)$ dictated by the numerical aperture (NA) of the measurement system. In the following, we will consider $V_{\mathrm{filt}}$, the reconstructible (i.e., filtered) potential, defined as follows:

\begin{align}
\label{E measured}
\tilde{V}_{\mathrm{filt}}(k_{x}, k_{y}, k_z) &=  W_{\mathrm{NA}}(k_x, k_y, k_z) \tilde{U}(k_x, k_y, k_z) \\
&=  W_{\mathrm{NA}}(k_x, k_y, k_z) \mathcal{F}_{3D}(VE)(k_x, k_y, k_z) \notag \\
\label{W filter}
W_{\mathrm{NA}}(k_x, k_y, k_z) &= H_{\mathrm{step}}(-k_z - n_\text{b}k_0\mathrm{NA}) \notag \\
&\quad \times \delta(\sqrt{k_x^2+k_y^2+k_z^2}-n_{\text{b}}k_0)
\end{align}
 where $H_{\mathrm{step}}$ is the Heaviside step function. $W_{\mathrm{NA}}$ is the transfer function of the imaging system. Its support is limited by the NA. It corresponds to the part of the Ewald sphere that can be grasped by the measurement system.  Note that we have here neglected the aberrations induced by the optical system.

\subsection{Information recovered under Born Series approximation}

In presence of multiple scattering, the source term $U=EV$ in \eqref{Helmholtz scattered} is too complex to be expressed directly. In order to demonstrate how a multiple scattering system can broaden the spatial filtering effect in comparison with the first-order Born hypothesis, we will make the second-order Born hypothesis. For this proof-of-concept, we will consider a specific type of medium containing a foreground object $V_1$ that we will intend to image and a background object $V_2$ that we will exploit to reach this goal.  We thus assume a medium of scattering potential $V = V_1 + V_2$ as described in Fig.~\ref{fig_1}(a).

\subsubsection{First-order Born transfer function}

Under Born's first-order assumption, there is no cross-interactions between the two objects due to the linearity of the model:
\begin{align}
\label{U first born}
U(x, y, z) &=  (V_1(x, y, z) + V_2(x, y, z))E_{\text{in}}(x,y,z) \notag \\
&=V_1(x, y, z)E_{\text{in}}(x, y, z) + V_2(x,y,z)E_{\text{in}}(x, y, z) \notag \\
&= U_1(x,y,z) + U_2(x,y,z)
\end{align}

Let us consider incident plane waves~\cite{Zhang2023,Lee2023,balondrade2024multi}:
$E_{\text{in}}(x, y, z) = e^{j(k_{\text{in},x}x+k_{\text{in},y}y+k_{\text{in},z}z)}$. We can then apply the convolution theorem to get~\cite{firstborn}:
\begin{align}
\label{U1 first born}
\tilde{U_{1}}(k_{x},k_{y},k_{z}) &= \tilde{V_{1}}(k_{x} -k_{\text{in},x},k_{y} - k_{\text{in},y},k_{z}- k_{\text{in},z})
\end{align}

Because of the reflection configuration $k_{\text{in},z}>0$ and only a part of the high $z$-frequencies of $V$ can be recovered by the measurements as illustrated in Fig.~\ref{fig_1}(c). 
This filtering phenomenon can nevertheless be mitigated by illuminating the sample with other plane waves at different angles. However, as illustrated in Fig.~\ref{fig_1}(d), this multi-illumination scheme still does not allow our system to recover the low spatial frequencies of the potential. 

Under the first Born approximation, only the potential's high $z-$frequencies can be recovered by the optical system. 
The problem is ill-posed in such a way that it is impossible to retrieve the average value of $V_1$, for instance. It therefore becomes mandatory to consider a multiple scattering model in order to recover the missing information.

\subsubsection{Second-order Born transfer function}

In order to exploit multiply-scattered waves and analytically evaluate their impact on the transfer function,
we assume in this subsection that $V_2(x, y, z) = \delta(z)v_2(x, y)$ and that the spatial support of $V_1$ is localized within a layer $[z_1, z_2]$, with $z_1 < z_2 < 0$. We choose $V_2$ as planar because we will consider it to be a reflector in our optimization problem, although in practice it never really is and it does not need to be a planar reflector as explained in detail in Section~\ref{subsecimportanceweightedloss}. We only need to be able to represent its reflectivity as the one of a plane reflector since this assumption allows us to express the second-order Born term analytically.

We model wave propagation by the second-order Born approximation which 
includes the contribution of double scattering events occurring within the medium:

\begin{align}
\label{U second born}
U(x, y, z) &= U_1(x,y,z) + U_2(x,y,z) + U_{11}(x,y,z) \notag \\ 
&+ U_{22}(x,y,z) + U_{12}(x,y,z) + U_{21}(x,y,z)
\end{align}

\noindent
where \( U_{ij} \) is defined as:
\begin{align}
\label{Uij}
U_{ij}(x, y, z) &= V_i(x,y,z)\iiint_{\mathbb{R}^3} G(x-x', y-y', z-z') \notag \\
&\quad \times V_j(x',y',z')E_{\text{in}}(x',y',z')\, dx'dy'dz'.
\end{align}

\begin{figure}[htbp]
    \includegraphics[width=1.0\linewidth]{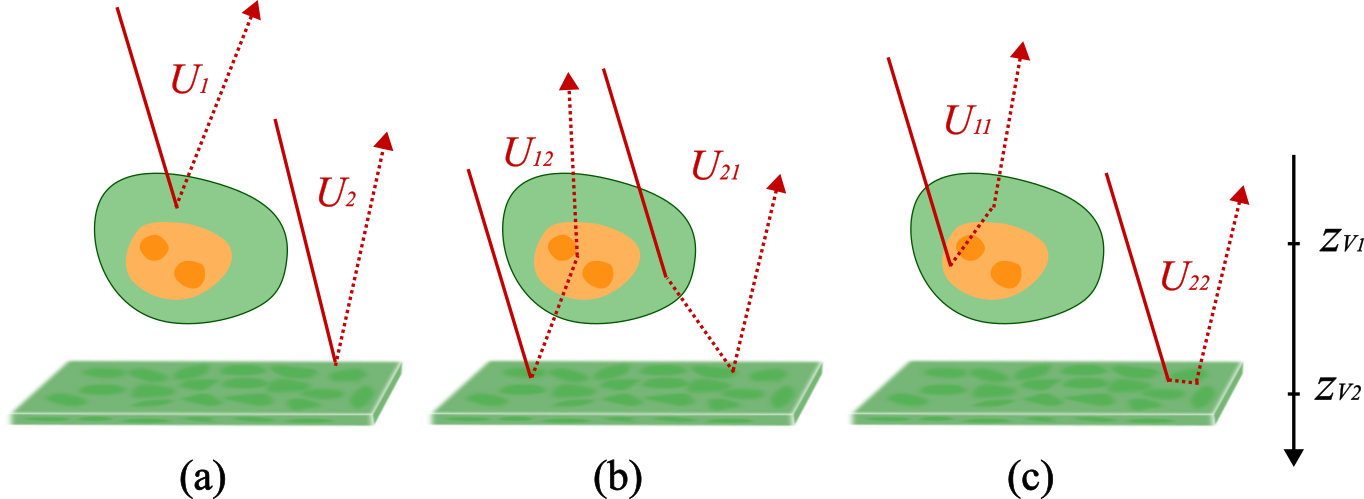}
    \caption{Decomposition of $U$ according to the different possible light scattering paths under second Born approximation.
(a) Representation of single scattering contributions.  
(b) Crossed terms $U_{12}$ and $U_{21}$ corresponding to successive interactions with the reflector and the target object.  
(c) Uncrossed double scattering contributions.
}
    \label{fig:fig8}
\end{figure}

We note that light scattering can here be expressed as the sum of three types of contributions: (a) single-scattering (or first-order Born) terms ($U_1$ and $U_2$), which encode the high spatial frequencies of the objects, (c) double-scattering terms ($U_{11}$ and $U_{22}$), corresponding to multiple scattering events occurring within each object and (b) cross-interaction terms ($U_{12}$ and $U_{21}$), which describe scattering paths involving a scattering event with respect to $V_1$ and a scattering event with respect to $V_2$. All these contributions are illustrated in Fig.~\ref{fig:fig8}.

In the following, we will demonstrate that cross-interaction terms encode the low $z$-frequencies of $V_1$. As the $V_2$-reflector is in the background relative to the $V_1$-medium, the $V_2$-reflector will act as a secondary source of waves propagating along the negative $z$-direction, 
which are “transmitted” waves from the observer's point of view. In the following, we will refer to this as indirect transmission, and our goal is to exploit this effect to reconstruct $V_1$. 

The calculation of these cross terms is detailed in Appendix \ref{appendix:B}. Assuming that the supports of $V_1$ and $V_2$ are far enough from each other to neglect evanescent waves, we obtain:
{\small
\begin{align}
\label{U12}
&\tilde{U}_{12}(k_x,k_y,k_z) = \iint_{\mathbb{R}^2} \frac{j \tilde{v}_{2}(k'_x-k_{\text{in},x}, k'_y-k_{\text{in},y})}{2\sqrt{n_{\text{b}}^2 k_{0}^{2} - k'_x{}^{2} - k'_y{}^2}} \\
& \times \tilde{V_1}(k_x - k'_x, k_y - k'_y, k_z + \sqrt{n_{\text{b}}^2 k_0^2-k'_x{}^2 - k'_y{}^2})\, dk'_x dk'_y \notag
\\
\label{U21}
&\tilde{U}_{21}(k_x,k_y,k_z)= \iint_{\mathbb{R}^2} \frac{j \tilde{v}_{2}(k_x-k'_x, k_y-k'_y)}{2\sqrt{n_{\text{b}}^2 k_{0}^{2} - k'_x{}^{2} - k'_y{}^2}} \\
& \times \tilde{V_1}(k'_x-k_{\text{in},x}, k'_y-k_{\text{in},y}, \sqrt{n_{\text{b}}^2 k_0^2-k_x^2 - k_y^2}-k_{\text{in},z})\, dk'_x dk'_y \notag
\end{align}
}
where $\tilde{v}_2$ is the 2D Fourier transform of $v_2$.
The two terms $\tilde{U}_{12}$ and $\tilde{U}_{21}$ are weighted convolutions between the reflector $\tilde{v}_2$ and the scattering potential $\tilde{V}_1$. These convolution products describe the fact that the 
presence of $v_2$ 
drastically enhances 
the spectral diversity of the field illuminating  
$V_1$. The amount of information about $V_1$ encoded in the components $U_{21}$ and $U_{12}$ is therefore increased compared to the first-order Born term $U_1$ \eqref{U1 first born}. In particular, the spectral diversity provided by $v_2$ allows the emergence of the low $z-$frequency components of $V_1$ in the measured wavefield  \eqref{E measured}. 
Thus, the information encoded in cross-interaction terms demonstrate that the measurement is sensitive to the low spatial frequencies of the object. A further analysis of the double-scattering contributions ($U_{11}$ and $U_{22}$ and higher other terms) is therefore unnecessary here, as our goal was only to show this dependence.

To make this dependence on low spatial frequencies more explicit, we consider specific reflector configurations that allow closed-form expressions for equations \eqref{U12} and \eqref{U21}. For instance, if the reflector is uniform (\textit{i.e.} $v_2(x,y)=v_{2o}$), we obtain:
\begin{align}
\label{v2 uniform}
&\tilde{U}_{12}(k_x,k_y,k_z) = \frac{jv_{2o}}{2\sqrt{n_{\text{b}}^2k_0^2-k_{\text{in},x}^2-k_{\text{in},y}^2}} \notag \\ 
& \times \tilde{V_1}(k_x-k_{\text{in},x}, k_y-k_{\text{in},y}, k_z + k_{\text{in},z})  
\\
&\tilde{U}_{21}(k_x,k_y,k_z) = \frac{jv_{2o}}{2\sqrt{n_{\text{b}}^2k_0^2-k_x^2-k_y^2}} \notag \\ 
& \times \tilde{V_1}(k_x-k_{\text{in},x}, k_y-k_{\text{in},y}, -\sqrt{n_{\text{b}}^2k_0^2-k_x^2-k_y^2} + k_{\text{in},z}) 
\end{align} 
This equation can also be found directly using the method of images \cite{taraldsen2005complex}. Applying equation \eqref{W filter}, it can be noted that encoded information $U_{21}$ is symmetrical with respect to the $k_z=0$ axis of the encoded information $U_{12}$, the filter of this configuration is presented in Fig.~\ref{fig_2a}. 

We can also consider the case of a sinusoidal reflector: $v_2(x,y) = v_{2o}\text{cos}(\gamma x)$ with Fourier transform $\tilde{v}_2(k_x,k_y) =\frac{v_{2o}}{2}(\delta(k_x-\gamma)+\delta(k_x+\gamma))$ with a normal incident wave $E_{\text{in}}(k_x,k_y,k_z)=e^{jn_{\text{b}}k_0z}$. For simplicity, the following expression is evaluated for $\tilde{v}_2(k_x,k_y)=v_{2o}\delta(k_x-\gamma)$ (Fig.~\ref{fig_2c}):
\begin{align}
\label{v2 sinuodal}
\tilde{U}&_{12}(k_x,k_y,k_z) = \frac{j v_{2o} \tilde{V_1}(k_x-\gamma, k_y, -k_z + \sqrt{n_{\text{b}}^2k_0^2-\gamma^2})}{2\sqrt{n_{\text{b}}^2k_0^2-\gamma^2}} \\ 
\tilde{U}&_{21}(k_x,k_y,k_z) = \frac{j v_{2o}}{2\sqrt{n_{\text{b}}^2k_0^2-(k_x-\gamma)^2-k_y^2}} \notag \\ 
& \times \tilde{V_1}(k_x-\gamma, k_y, \sqrt{n_{\text{b}}^2k_0^2-(k_x-\gamma)^2 - k_y^2}-n_{\text{b}}k_0)
\end{align}

In the following section, we use these two peculiar configurations to illustrate how the background reflector $V_2$ can be leveraged to retrieve the low $z-$frequencies of the foreground structure $V_1$.

\subsection{Filtering effect for particular configurations}

To that aim, we describe the inverse problem of light scattering as the following optimization problem:
\begin{align}
\label{optim}
\hat{\mathbf{V}} &=  \mathop{\arg\min}\limits_{\mathbf{V} \in \mathbb{R}^{N}} \mathcal{L}(\mathbf{V}) \\
\mathcal{L}(\mathbf{V}) &=  \| \mathbf{y}_{\text{model}}(\mathbf{V}, \mathbf{E}_{\text{in}}) -\mathbf{E}_{\text{m}}\|^2
\end{align}

Here, $\mathbf{V}$ and $\mathbf{E}_{\text{in}}$ represent discrete approximations of the potential $V$ and the incident field $E_{\text{in}}$ described respectively as vectors in $\mathbb{R}^{N}$ and $\mathbb{C}^{N}$ (the square domain is discretized as a uniform grid of size $N = N_x \times N_y$). The objective is to minimize a loss function over one illumination to obtain an estimate of the potential $V$ consistent with equations \eqref{v2 uniform} and \eqref{v2 sinuodal}. Equation \eqref{optim} represents the fidelity term for a single measurement, quantifying the similarity between the field predicted by the forward model, $\mathbf{y}_{\text{model}}$, and the measured scattered field $\mathbf{E}_{\text{m}}$.

\begin{figure}[htbp]
    \centering
    \captionsetup[subfloat]{font=scriptsize}
    \subfloat[]{%
        \includegraphics[width=0.22\textwidth]{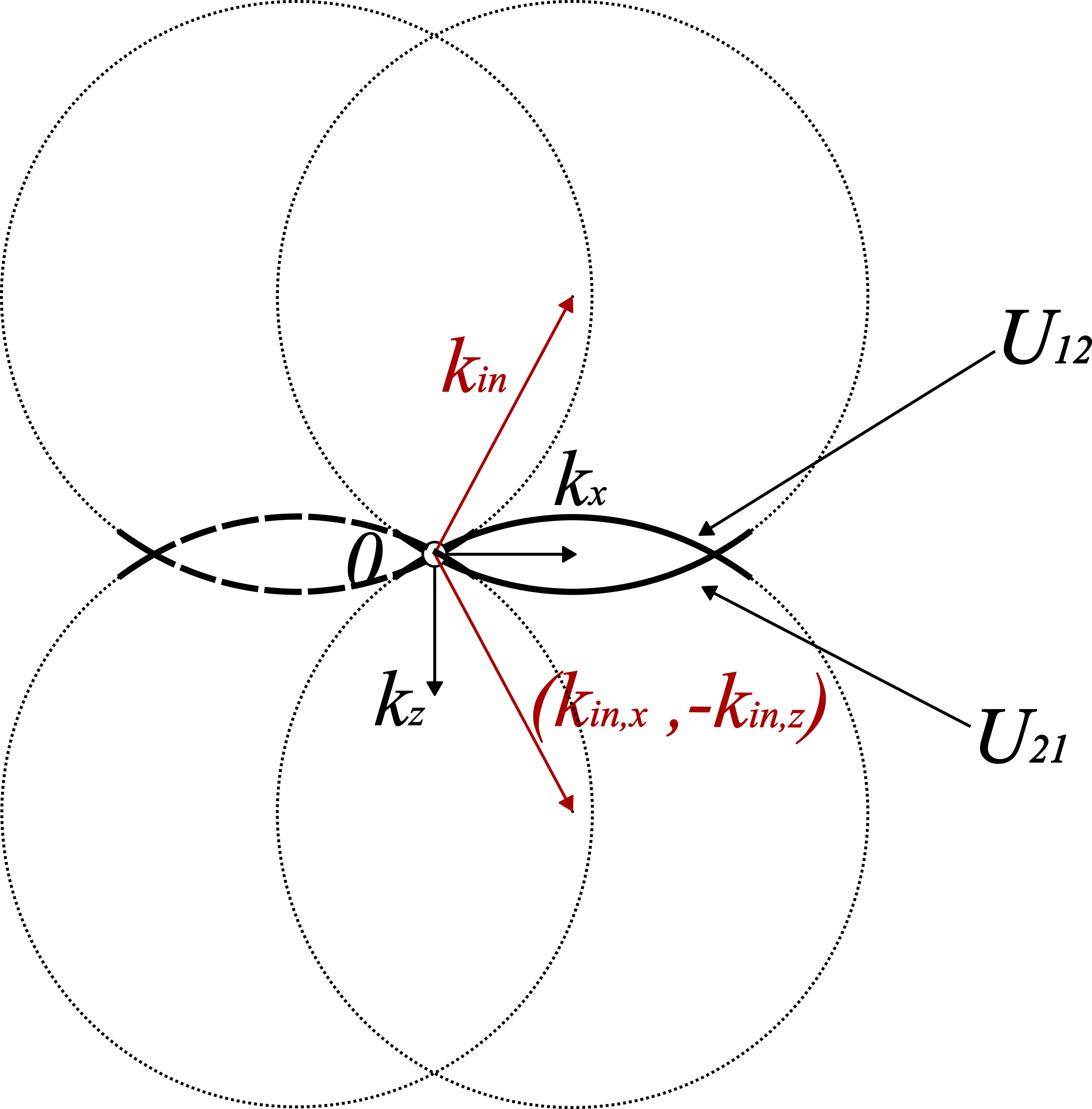}%
        \label{fig_2a}
    }\hfill
    \subfloat[]{%
        \includegraphics[width=0.22\textwidth]{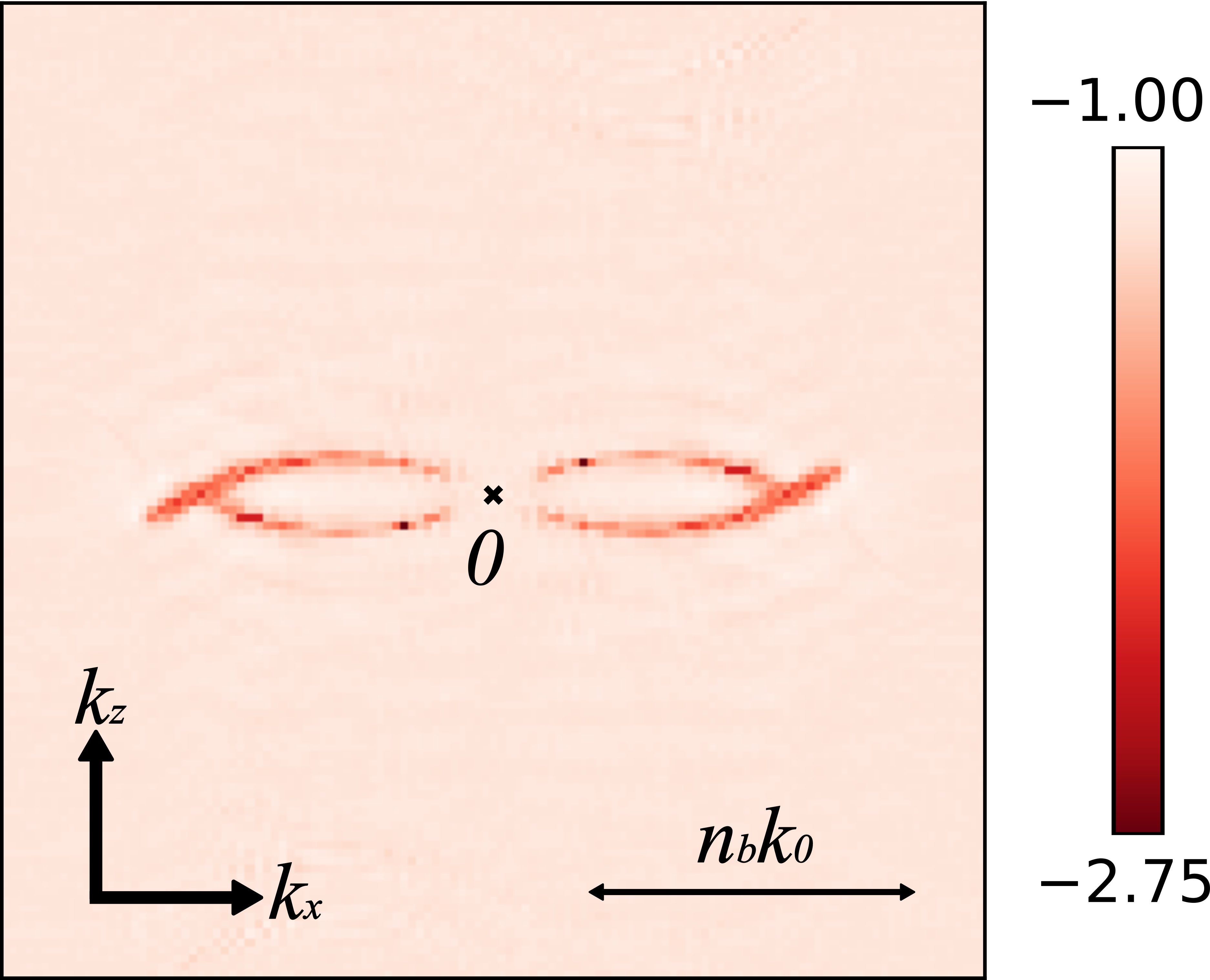}%
        \label{fig_2b}
    }\\ 

    \subfloat[]{%
        \includegraphics[width=0.22\textwidth]{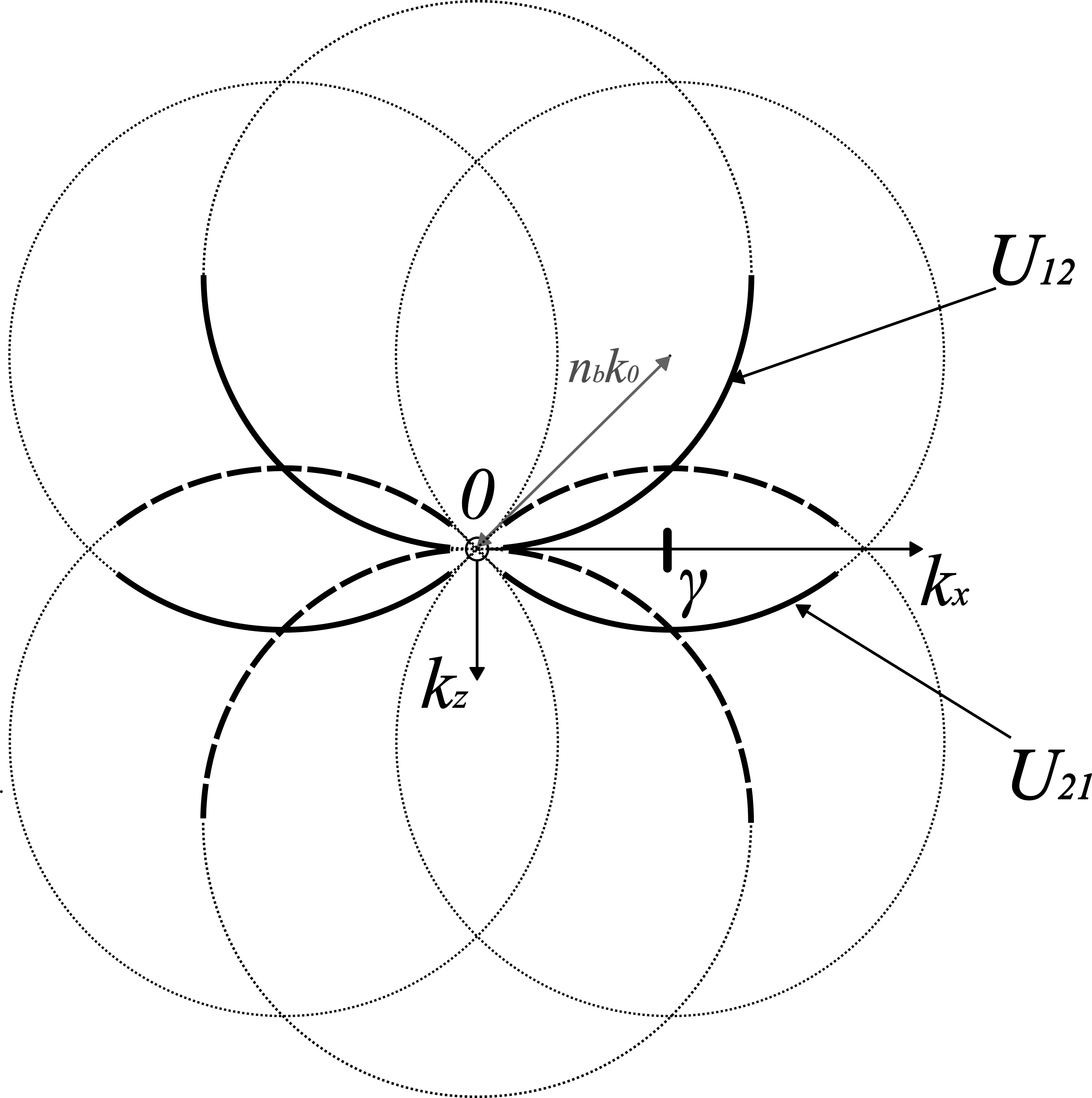}%
        \label{fig_2c}
    }\hfill
    \subfloat[]{%
        \includegraphics[width=0.22\textwidth]{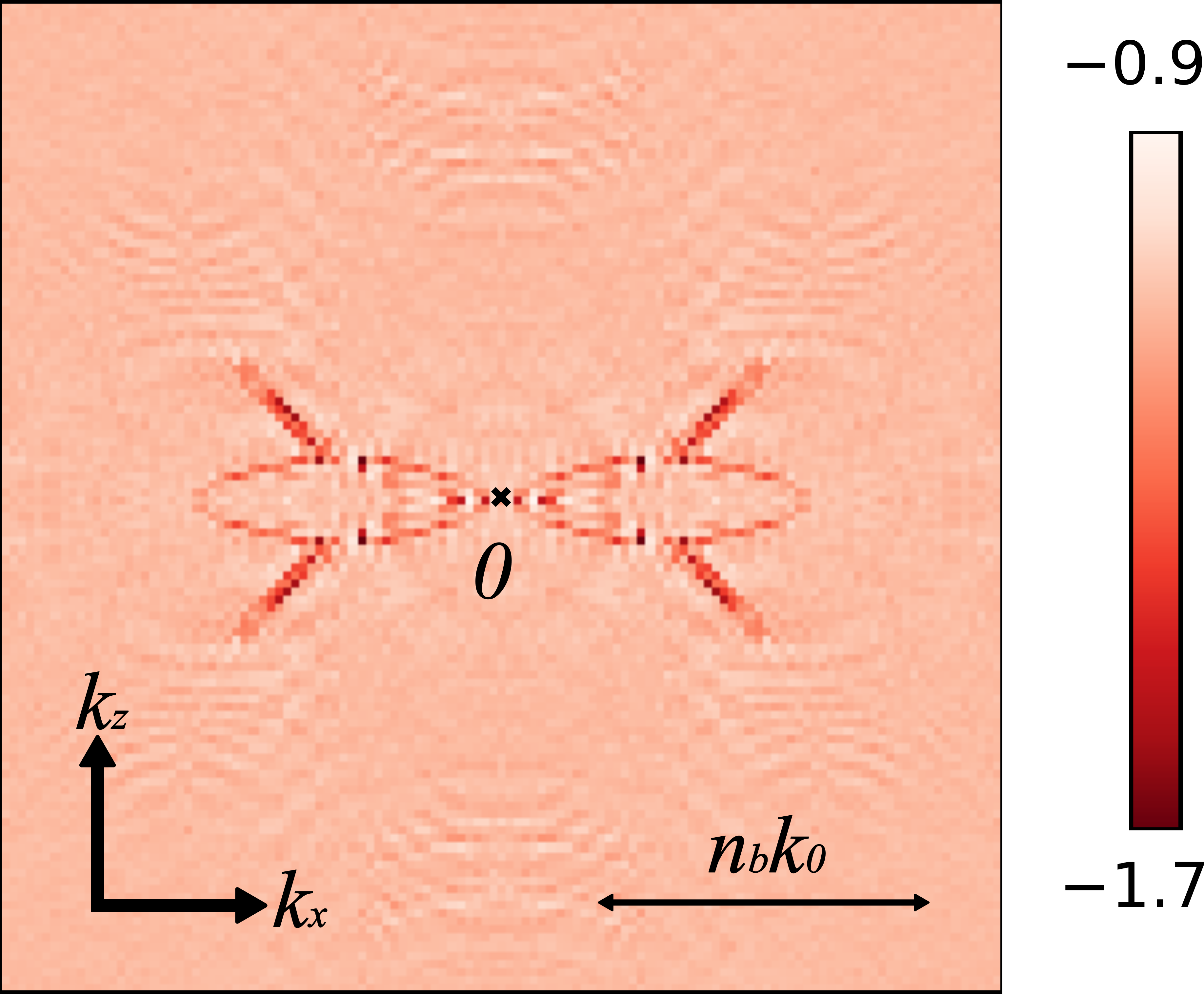}%
        \label{fig_2d}
    }

    \caption{Support of the spatial frequency spectrum of the object $V_1$ extracted thanks to the background reflector $V_2$.
        (a) Sketch of these spectral components for a uniform reflector \eqref{v2 uniform} and an incident plane wave of wave vector $\mathbf{k}_{\textrm{in}}=(k_{\text{in},x}, k_{\text{in},z})$. (b) Reconstruction error in the spatial Fourier domain between the ground truth $V_1$ and its estimator $\hat{V}_1$ deduced from the optimization process in 2D \eqref{optim} (with red indicating better results). (c) Sketch of the accessible spectral components of $V_1$ for a cosine-like reflector of frequency \(\gamma\) \eqref{v2 sinuodal} and a normal incident plane wave ($k_{\text{in},x} = 0$). (d) Same as in panel (b) but for the cosine-like reflector considered in panel (c). In panels (a) and (c), the dashed lines are deduced from the real-value character of $V_1$ which implies a symmetric spectrum in the Fourier domain. Panels (b) and (d) are numerically obtained in 2D by considering a Dirac potential $V_1$ whose Fourier spectrum is uniform.
}
    \label{fig_2}
\end{figure}

As a forward model, we will use the second order Born in order to take into account the predominant multiple scattering paths.
The computed total field becomes: 
\begin{align}
\label{secondbornforward}
\mathbf{E}_{\text{Born-2}}(\mathbf{V}, \mathbf{E}_{\text{in}}) &= \mathbf{E}_{\text{in}} + \mathbf{G} \text{diag}(\mathbf{V})\mathbf{E}_{\text{in}} \notag \\ 
&+ \mathbf{G} \text{diag}(\mathbf{V})\mathbf{G} \text{diag}(\mathbf{V})\mathbf{E}_{\text{in}} \\
\textbf{y}_{\text{Born-2}}(\mathbf{V}, \mathbf{E}_{\text{in}}) &= \mathbf{H}\text{diag}(\mathbf{V})\mathbf{E}_{\text{Born-2}}(\mathbf{V}, \mathbf{E}_{\text{in}})
\end{align}
where $\mathbf{H} \in \mathbb{C}^{N_{\text{c}} \times N}$ is the transfer function describing wave propagation from the sample to the $N_c$ pixels of the measurement system and $\text{diag}(\mathbf{V}) \in \mathbb{C}^{N \times N}$ the square diagonal matrix whose diagonal elements are given by the components of $\mathbf{V}$.
In real space, this model involves convolution (or matrix) products, which are computationally expensive. However, this can be bypassed by moving to Fourier space and using Fourier transforms of Green's functions: $\mathbf{G} = \mathbf{F}^{\dagger}\text{diag}(\tilde{\mathbf{G}})\mathbf{F} \in \mathbb{C}^{N \times N}$. Here, $\mathbf{F} \in \mathbb{C}^{N \times N}$ is the Fourier transform operator, which is efficiently computed in practice using the Fast Fourier Transform (FFT) algorithm. The computational cost of the forward model is then reduced
as it only involves a sequence of term-by-term products and FFT operations.
Because of the singularity of the Green's function in Fourier space, we use modified versions $\tilde{\mathbf{G}}^L \in \mathbb{C}^{N}$, as defined in Refs.~\cite{soubies2017efficient, pham2020three, vico2016fast} to ensure computation stability. This modified Green's function correspond to cropped version in real space, and a twofold zero-padding is required to ensure accurate computation.

In the following, we will attempt to solve \eqref{optim} using the gradient descent algorithm. However, the forward model \eqref{secondbornforward} is a second-order polynomial function of $V$, and gradient descent does not have theoretical guarantees to converge to a global solution of \eqref{optim}. To circumvent this problem and to illustrate the information that can be recovered from the system without dealing with convergence issues (which will be addressed in the next sections), we will fix the background reflector in the simulation as the ground truth.

 The model used to calculate the ground truth field is the Lippmann-Schwinger forward model described in Ref.~\cite{soubies2017efficient}. For sake of illustration of the filtering effect operated by our imaging method, the scattering potential $V_1$ that we consider here is a Dirac point, as it gives rise to a uniform spectrum in Fourier space. Figures~\ref{fig_2}(b) and (d) show the reconstruction error between the estimated potential $\hat{V}_1$ and its ground truth $V_1$ in Fourier space for the uniform and sinusoidal reflectors $V_2$. A quantitative agreement is found between $\hat{V}_1$ and $V_1$ over the spatial frequencies captured by the imaging system. The recovered spatial frequency spectra of ${V}$ nicely match with our analytical predictions \eqref{v2 uniform} and \eqref{v2 sinuodal} sketched in Figs.~\ref{fig_2}(a) and (c), respectively. In particular, it illustrates how the presence of reflector $V_2$ enables multiply scattered trajectories that can be leveraged for retrieving the low $z-$frequency components of the foreground object $V_1$.

\section{Reconstruction algorithm}

As discussed in the previous section, 
an internal background reflector helps to retrieve the low-frequency information from the object in the foreground. However, this information can only be obtained by analyzing the multiply scattered components of the reflected wavefield
because the single scattering contribution does not contain the low spatial frequencies of the foreground object. As a result, for reflection-mode ODT, the forward model cannot be linear with respect to the scattering potential, making the optimization problem non-convex. Gradient descent algorithms, such as stochastic gradient descent (SGD) or even FISTA \cite{fista}, therefore lack theoretical guarantees for convergence. 

In this section, we propose a reconstruction algorithm that enables good reconstruction in reflection configurations despite the non-convexity of the problem. Our approach is based on (i) the introduction of a new loss function that focuses on the low-frequency components of the foreground object to be reconstructed, (ii) the use of Total Variation (TV) regularization (a commonly applied method \cite{kamilovtv, RUDINTV} that constrains the reconstruction to block-wise behavior) in the foreground region only (i.e. between $z_1$ and $z_2$ in Fig.~\ref{fig_1}), and (iii) a proximal algorithm based on the Adam optimizer \cite{proxgen} ensuring a good convergence despite the non-convexity of the problem.

\subsection{A temporal loss function to focus on low-frequency information}

As illustrated in equation \eqref{U second born}, the total information recovered is a sum of terms which encode different pieces of information: $U_{1}$ encodes high $z-$frequencies of the object $V_1$ \eqref{U1 first born} while $U_{12} $ \eqref{U12} and $ U_{21}$ \eqref{U21} encode its low $z-$frequencies. In the ideal case, if $V_2$ is known, then we can reconstruct both low and high frequencies of the object. However, in practice, the orders of magnitudes of these two terms can be very different, leading to difficulties for the reconstruction of the low $z-$frequencies of the foreground object if $V_2$ is too small. We could assume that $V_2$ is on average much larger than $V_1$, but this is not necessarily the case in reality. Ideally, the terms related to direct reflection with the object should be separated from those containing cross interactions between the reflector and the object to be imaged. This can be done by considering the reflected wavefield in the time domain, as it can be done in ultrasound with broadband signals~\cite{Weber2021}.

In optical microscopy, the time-dependence of the reflected wavefield can be retrieved by low coherence interferometry~\cite{Badon2015}.
An alternative strategy is to perform a polychromatic measurement of the wavefield using a spectrometer~\cite{Choma2003,Yun_03} or a swept source laser~\cite{hillmann_efficient_2012,povazay_full-field_2006} as in spectral domain OCT.
In this way, $\textbf{E}_{\text{m}}$ can be measured at different temporal frequencies 
$\omega$.
In post-processing, an inverse Fourier transform with respect to $\omega$ can then provide the time-dependent wavefield $\textbf{E}_{\text{m}}(t) $ that would be  obtained  if  the  sample  was  illuminated by a coherent incident wave-packet.

With $V_1$ in the foreground and $V_2$ in the background, the ballistic term $U_1$ and crossed terms $U_{12}$ and $U_{21}$ can be associated with different central times-of-flight $t_1 \sim 2 n_{\text{b}} z_{V_1}/c$ and $t_2 \sim 2 n_{\text{b}} z_{V_2}/c$, with $z_{V_1}$ and $z_{V_2}$, the central depths of the foreground object $V_1$ and the background reflector $V_2$ (Fig.~\ref{fig:fig8}). This assumption is valid in a weakly scattering regime in which the echo time $t$ is directly proportional to the depth $z$ of the associated reflector. Otherwise, high-order multiple scattering would lead to a large temporal dispersion of the incident wave packet~\cite{Derode1995,Mosk2012,montecarlo}, making their discrimination impossible on a time-of-flight basis.

By considering the $i^{\text{th}}$ Born order forward model, we have: 
\begin{align}
\textbf{E}_{\text{m}}(t) &= \mathcal{F}_{\omega}^{-1}[\textbf{E}_{\text{m}}(\omega)] (t)\\
\textbf{y}_{\text{Born-}i}(\mathbf{V}, \mathbf{E}_{\text{in}}, t) &= \mathcal{F}_{\omega}^{-1}[\textbf{y}_{\text{Born-}i}(\mathbf{V}, \mathbf{E}_{\text{in}}, \omega)] (t)
\end{align}
We then define a new temporal loss function over the $N_{\text{in}}$ incident illuminations~\cite{balondrade2024multi}:
\begin{align}
\label{temporal loss} 
\mathcal{L}(\mathbf{V}, t) &= \sum_{j=1}^{N_{\text{in}}} \mathcal{L}_j(\mathbf{V}, t)\\ \mathcal{L}_j(\mathbf{V}, t) &=  \| \mathbf{y}_{\text{Born-}i}(\mathbf{V}, \mathbf{E}_{\text{in},j}, t) -\mathbf{E}_{\text{m},j}(t) \|^2  ,
\end{align}
and the fidelity term of the loss function becomes
\begin{align}
\label{reflection loss} 
\mathcal{L}_{\text{fid}}(\mathbf{V}, \gamma) = \mathcal{L}(\mathbf{V}, t_2) + \gamma\mathcal{L}(\mathbf{V}, t_1)
\end{align}
The purpose of this loss function is to treat the information from direct reflection of the foreground object (associated with time $t_1$) and the reflection from the background reflector and the cross interaction terms (associated with time $t_2$) with different weights.
 If the background reflector has a low reflectivity and $\gamma =1$, then the minimization algorithm of the loss would prioritize the reconstruction of the high spatial frequencies of the foreground object, which may cause the whole optimization process to become inefficient.
When $\gamma$ is small, the impact of the direct reflection from the object is reduced. As a result, the information about the object to be imaged becomes a nonlinear function of its spatial low frequencies, effectively "forcing" the model to focus on a solution depending on the object's low-frequency components.

Thus, by considering the problem from a temporal perspective, we can weight the two effects. It becomes even possible, by adjusting $\gamma$, to recover both high and low spatial frequencies of the foreground object, improving the reconstruction quality of the sample.  We will explain in Appendix C how $\gamma$ is selected. Moreover, $t_1$ and $t_2$ can be considered as specific moments or as time intervals. In practice, the use of time intervals leads to better reconstruction results.

It is important to note that minimizing \eqref{reflection loss} remains a non-convex problem, which can be challenging even for classical gradient descent algorithms.
We address this issue in the next section.

\subsection{Regularization and optimization algorithm}

The non-convexity of the loss function can lead to multiple local minima or saddle points in the loss landscape in \eqref{reflection loss}. Using a regularization term introduces \textit{a priori} information that helps reduce the number of possible solutions and improves the convergence of gradient-based methods.
The objective function to minimize in our problem becomes:
\begin{align}
\label{rODT loss} 
\mathcal{L}_{\text{rODT}}(\mathbf{V},\gamma,\tau) = \mathcal{L}(\mathbf{V}, t_2) + \gamma\mathcal{L}(\mathbf{V}, t_1) + \tau\mathcal{R}(\mathbf{V})
\end{align}
in which $\mathcal{R}$ denotes the regularization term and $\tau$ is the regularization parameter.

A commonly used regularization term is the Total Variation loss (TV) \cite{RUDINTV}, which is today commonly used for ODT in transmission configuration \cite{kamilovtv, chen2020multi}. This regularization encourages the reconstruction of piecewise constant potentials and avoid noisy reconstructions. Noise can arise, for example, if the reflector is not smooth, making the reconstruction task more difficult.
A widely used optimization approach for ODT in transmission configuration is proximal algorithms \cite{parikh2014proximal}, which are well suited for TV regularization by allowing the iterative optimization of loss functions composed of differentiable and non-differentiable terms. Let $\mathbf{V}^{(k)}$ be the estimator of $\mathbf{V}$ at iteration $k$. The idea is to split each optimization step into two parts:
\begin{itemize}
\item[(i)] A standard gradient descent step is first performed: $$\mathbf{V}^{(k)}\gets\mathbf{V}^{(k)}-\alpha\nabla \mathcal{L}_{\text{fid}}(\mathbf{V}^{(k)},\gamma) . $$
\item[(ii)]The proximal operator is then applied: $$\mathbf{V}^{(k+1)}\gets\text{prox}_{\alpha \tau \mathcal{R} }(\mathbf{V}^{(k)}),$$ which corresponds to the minimization of:
\begin{align}
\hspace*{-0.25in} \text{prox}_{\alpha \tau \mathcal{R} }(\mathbf{V}^{(k)}) = \mathop{\arg\min}\limits_{\mathbf{V} \in \mathbb{R}^{N}} \left\{ \frac{1}{2}\|\mathbf{V}^{(k)} -\mathbf{V}\|^2 + \alpha\tau\mathcal{R}(\mathbf{V}) \right\} \notag
\end{align}
\end{itemize}

The proximal operator for TV regularization has the advantage of being well-documented \cite{proximaltv, kamilovtv}. 

FISTA and variants are gradient descent or stochastic gradient descent based algorithms. This class of algorithms are designed for convergence of convex loss functions. Reflection ODT being a non convex problem, it was natural to search for a proximal algorithm based on an optimizer efficient for non-convex optimization.
In that respect,  an adequate tool is the Adam optimizer~\cite{kingma2014adam} which is an adaptive learning rate method. This optimizer is suitable for stochastic gradient descent which allows a reduction of the memory requirement for the optimization process. Moreover, Adam showed better efficiency in non-convex optimization compared to Nestervov stochastic gradient descent algorithm \cite{de2018convergence, zaheer2018adaptive}. 

The challenge with adaptive algorithms lies in the fact that the learning rate depends on the parameter being optimized, thereby requiring a proximal operator that accounts for this dependency. 
Ref.~\cite{proxgen} introduces a method called \textit{ProxGen} that extends proximal gradient algorithms to adaptive optimization techniques such as Adam. Based on this approach, we will use the following proximal algorithm in our methodology:

\begin{algorithm}[H]
\caption{Reflection ODT algorithm: \textit{ProxGen}}\label{alg:alg1}
\begin{algorithmic}
\STATE 
\STATE {\textbf{input :}} $\mathbf{V}^{(0)}$, $\alpha$, $\epsilon$, $\beta_1$, $\beta_2$, $\gamma$ and $\tau$

\STATE Initialize $\mathbf{m}_0=0$ and $\mathbf{v}_0=0$

\STATE {\textbf{for $k = 1,2,...,N_{\text{epoch}}$ do:}} 
\STATE \hspace{0.3cm}{Choose randomly an index $i_{k} \in \left\{ 0, ...,N_{\text{in}} \right\} $}
\STATE \hspace{0.3cm}$ \mathbf{g}_{k} \gets \nabla \mathcal{L}_{\text{fid}, i_{k}}(\mathbf{V}^{(k-1)},\gamma)$
\STATE \hspace{0.3cm}$ \mathbf{m}_{k} \gets \beta_{1}\mathbf{m}_{k-1} + (1-\beta_{1})\mathbf{g}_{k}$
\STATE \hspace{0.3cm}$ \mathbf{v}_{k} \gets \beta_2\mathbf{v}_{k-1} + (1-\beta_2)(\mathbf{g}_{k}\circ \mathbf{g}_{k})$
\STATE \hspace{0.3cm}$ \mathbf{V}^{(k)} \gets \mathop{\arg\min}_{\mathbf{V} \in \mathbb{R}^{N}} \lbrace \langle \mathbf{V},\mathbf{m}_k \rangle + \tau\mathcal{R}(\mathbf{V})$
\STATE \hspace{5.05cm}$+\frac{\|\mathbf{V}-\mathbf{V}^{(k-1)}\|^2_{ {\sqrt{\mathbf{v}_k}}+\epsilon}}{2\alpha} \rbrace$

\textbf{return}  The prediction $\mathbf{V}^{(N)}$
\end{algorithmic}
\label{alg1}
\end{algorithm}

In the algorithm we have introduced a new norm as follows:  
$\|\mathbf{x}\|^2_{\mathbf{a}} = \langle \mathbf{x},\mathbf{a} \circ \mathbf{x} \rangle$
where $\circ$ denotes the element-wise (Hadamard) product. 
The hyperparameters of the algorithm are: the learning rate $\alpha$, the momentum parameters $\beta_1$ and $\beta_2$, the weight factor $\gamma$, and the regularization parameter $\tau$.
The last line of the algorithm corresponds to a minimization problem, which can be reformulated using a new proximal operator:
\begin{align}
\label{reflction loss} 
\mathbf{V}^{(k)} &\in \mathop{\arg\min}\limits_{\mathbf{V} \in \mathbb{R}^{N}} \left\{\langle \mathbf{V},\mathbf{m}_k \rangle + \tau\mathcal{R}(\mathbf{V}) +\frac{\|\mathbf{V}-\mathbf{V}^{(k-1)}\|^2_{\sqrt{\mathbf{v}_k}+\epsilon}}{2\alpha} \right\} \notag \\
&= \text{prox}_{\alpha \tau \mathcal{R} }^{\sqrt{\mathbf{v}_k}+\epsilon} \left (\mathbf{V}^{(k-1)}-\frac{\alpha}{\sqrt{\mathbf{v}_k}+\epsilon}\mathbf{m}_k \right )
\end{align}
with $\text{prox}_{h}^{\mathbf{a}}(\mathbf{z}) = \arg\min\limits_{\mathbf{x}} \left\{ h(\mathbf{x}) + \frac{1}{2} \|\mathbf{x}-\mathbf{z}\|^2_{\mathbf{a}}  \right\}$ and the  division is performed element-wise.
This minimization problem is convex and can be solved by the proximal algorithm \cite{melchior2020proximaladamrobustadaptive}.
The gradient of the objective function is $(\sqrt{\mathbf{v}_k}+\epsilon)(\mathbf{x}-\mathbf{z})$ and the Lipschitz constant is $L_k = {\text{max}(\sqrt{\mathbf{v}_k}+\epsilon)}$.

The algorithm to compute efficiently $\mathbf{V}^{(k)}$ is then:

\begin{algorithm}[H]
\caption{Proximal calculus}\label{alg:alg2}
\begin{algorithmic}
\STATE 
\STATE {\textbf{input :}} $\mathbf{V}^{(k-1)}$, $\alpha$, $\epsilon$, $\mathbf{m}_k$, $\mathbf{v}_k$ and $\tau$
\STATE $\mathbf{V}_{\text{prox}}^{(k,0)} = \mathbf{V}^{(k-1)}$
\STATE $\mathbf{z}_k = \mathbf{V}^{(k-1)}-\frac{\alpha}{\sqrt{\mathbf{v}_k}+\epsilon}\mathbf{m}_k$
\STATE {\textbf{for $t = 1,2,..., T$ do:}} 
\STATE \hspace{0.3cm}$ \mathbf{g}_{t} \gets (\sqrt{\mathbf{v}_k}+\epsilon)(\mathbf{V}_{\text{prox}}^{(k,t-1)}-\mathbf{z}_k)$
\STATE \hspace{0.3cm}%
$\mathbf{V}_{\text{prox}}^{(k,t)} \gets 
\text{prox}_{\tau \alpha \!/\!  \max(\sqrt{\mathbf{v}_k}+\epsilon)\mathcal{R}}
(\mathbf{V}_{\text{prox}}^{(k,t-1)} \!-\! 
\frac{\mathbf{g}_t}{\max(\sqrt{\mathbf{v}_k}+\epsilon)})$

\STATE \textbf{return}  $\mathbf{V}^{(k)} =\mathbf{V}_{\text{prox}}^{(k,T)}$
\end{algorithmic}
\label{alg2}
\end{algorithm}

The parameter $T$ defines the number of iterations for Algorithm 2, thereby controlling its convergence behavior. In practice, we set $T$ to be large and apply a stopping criterion which corresponds to the relative change from one iteration to the next:

\begin{align}
\label{stopping criterion} 
\frac{\| \mathbf{V}_{\text{prox}}^{(k,t)} - \mathbf{V}_{\text{prox}}^{(k,t-1)} \|}{\| \mathbf{V}_{\text{prox}}^{(k,t)}\|} \leq \delta
\end{align}
with $\delta = 10^{-6}$ in all following simulations.

\subsection{Simulation results}

To validate our reconstruction algorithm, we applied it to 2D simulations for computational efficiency. 3D-simulated medium results are presented at the end of the section. For both 2D and 3D cases, to ensure consistency with the theoretical framework---which assumes the convergence of the Born series---the RI deviations are set to $\Delta n = 0.01$. This RI contrast, although not representative of all biological media, is comparable to the RI contrasts of biological samples considered in recent work \cite{Chowdhury:19,app12030951,10.3389/fphy.2021.666256}. The surrounding medium has a RI $n_{\text{b}} = 1.33$.
We consider incident plane waves with angles confined within a numerical aperture of $\text{NA} = 0.8$, and wavelengths ranging from 800 to 875 nm.

For both 2D and 3D, the ground truth is, as above, generated using a Lippmann-Schwinger forward model with a modified Green’s function $\tilde{\mathbf{G}}^L$ \cite{pham2020three, vico2016fast}. This model accurately accounts for all multiple-scattering effects and is a well-established method for simulating light propagation in complex media. The use of these modified Green’s functions requires adding a padding region with a RI $n_{\text{b}}$, whose length matches that of the sample. The gradient for our reconstruction is calculated by using the autograd functionality of torch library \cite{paszke2017automatic}. For each reconstruction, the reflector is not known in advance and the starting RI map is always a uniform medium with a RI equal to $n_{\text{b}}$. \\

\subsubsection{2D simulation}
\begin{figure}[htbp]
    \centering
    \includegraphics[width=1.0\linewidth]{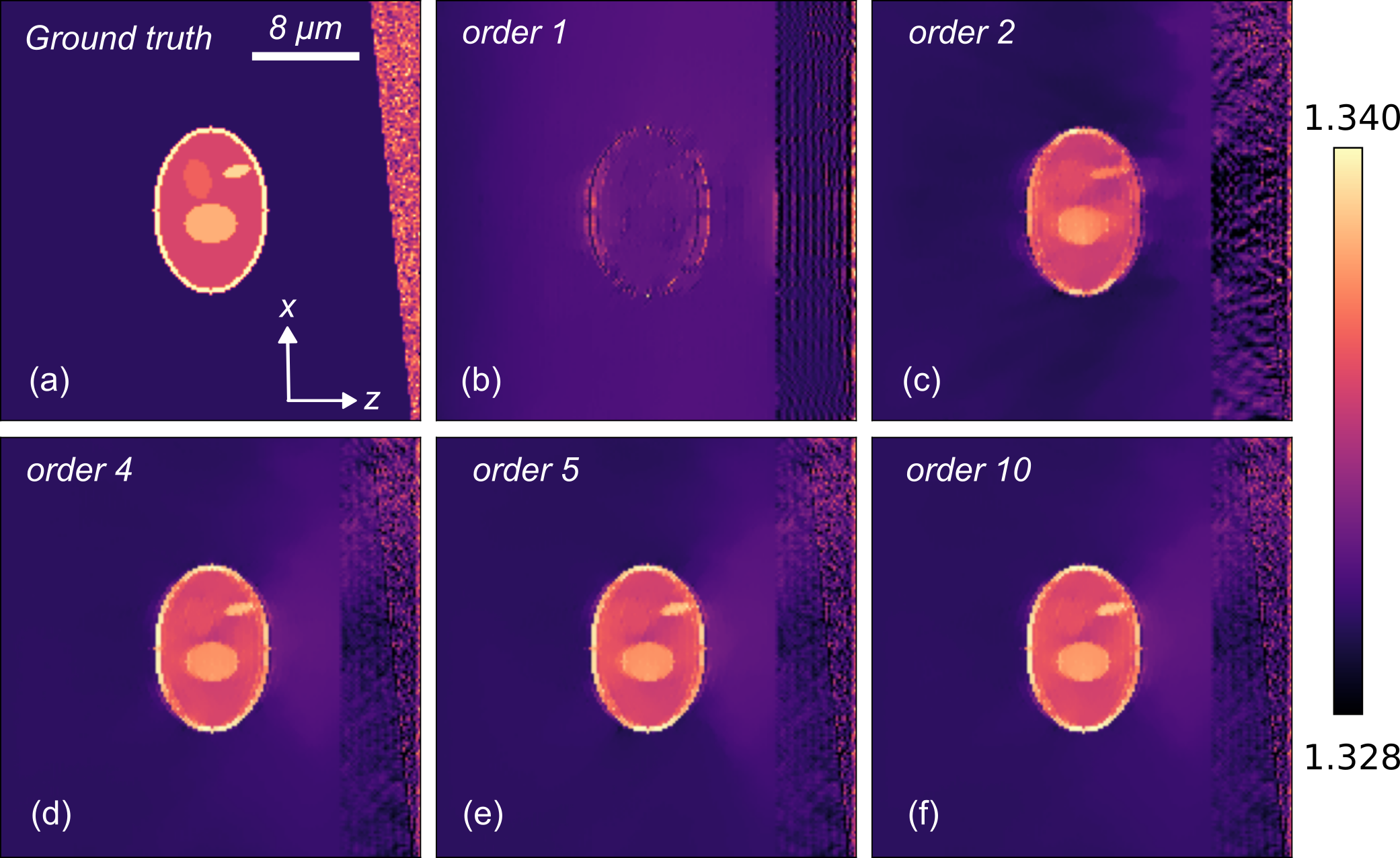}
    \caption{Reconstruction of a sample composed of a cell and a reflector for different forward models used in the  \textit{ProxGen} algorithm with the weighted temporal loss function ($\gamma=0.1$). A minimum regularizer is used to improve the quality of reconstruction. The reconstruction is impossible by using first order Born approximation.}
    
    \label{fig:fig4}
\end{figure}

To evaluate the effectiveness of our approach, we tested our reconstruction algorithm on 2D simulated samples with a resolution of $dx = dz = 200$ nm in a cube of size $N_x = N_z = 161$. Samples under study mimic the RI distribution of a cell. We have taken $N_{\text{in}}=80$ incident waves taken uniformly and sampled over 11 wavelengths to guarantee a sufficient temporal resolution and discriminate the direct object echo and the background reflector components.
The reflector placed in our simulation behind the sample corresponds to an interface with average RI $n=1.335$, to which we add Gaussian noise with standard deviation $\delta n =0.002$.
For optimal convergence, all 2D hyperparameters have been carefully tuned for Figs. \ref{fig:fig4}-\ref{fig:fig5}. We also validated our approach using  simulated  noisy  observations.  Further  details  can be found in Appendix C. 

\subsubsection{Choice of the forward model}
\label{subsubsec:choiceforwardmodel}%
Choosing an appropriate forward model for the optimization algorithm is essential for improving the quality of the RI reconstruction. Indeed, this model should closely account for the multiple-scattering contribution in order to accurately reconstruct the sample. Fig. \ref{fig:fig4} illustrates this fact by displaying the result of our optimization algorithm for different forward models. Each model involves a different order of the Born approximation. Not surprisingly, the first-order Born model (Fig. \ref{fig:fig4}(b)) only grasps the high spatial frequencies of the sample and only provides an image of the external surface of the cell. Remarkably, the second-order Born model reveals the intra-cellular components (Fig. \ref{fig:fig4}(c)). Nevertheless, the reconstruction remains imperfect due to discrepancies between the real light propagation and the forward model used in the optimization algorithm. Those imperfections can be mitigated by increasing the degree of complexity of the forward model (Fig. \ref{fig:fig4}(d)-(f)). The 5$^{\text{th}}$-order Born series forward model (Fig. \ref{fig:fig4}(e)) offers a good trade-off between the accuracy of the RI map reconstruction and the computational complexity. This forward model includes more scattering events than in our theoretical analysis. One might think that the reconstruction quality also depends on other multiple-scattering terms, not just the cross-interactions $U_{12}$ and $U_{21}$. Yet this effect remains minor compared to the contribution of the cross-interaction terms $U_{12}$ and $U_{21}$, as detailed in Appendix~\ref{appendix:D}.

\subsubsection{Regularization strategies}

In the following, total variation (TV) regularization is applied only in the region of the object $V_1$ that we aim to reconstruct. Since the reflector is difficult to reconstruct due to the ill-posed nature of the problem, we choose not to apply any regularization or constraint to it. This allows the reconstruction to produce a RI that, while not physically accurate, effectively mimics the reflector's reflectivity. Applying regularization only in foreground also improves the reconstruction, as shown in Appendix \ref{appendix:C}. 

\begin{figure*}[t]
    \centering
    \includegraphics[width=1.0\linewidth]{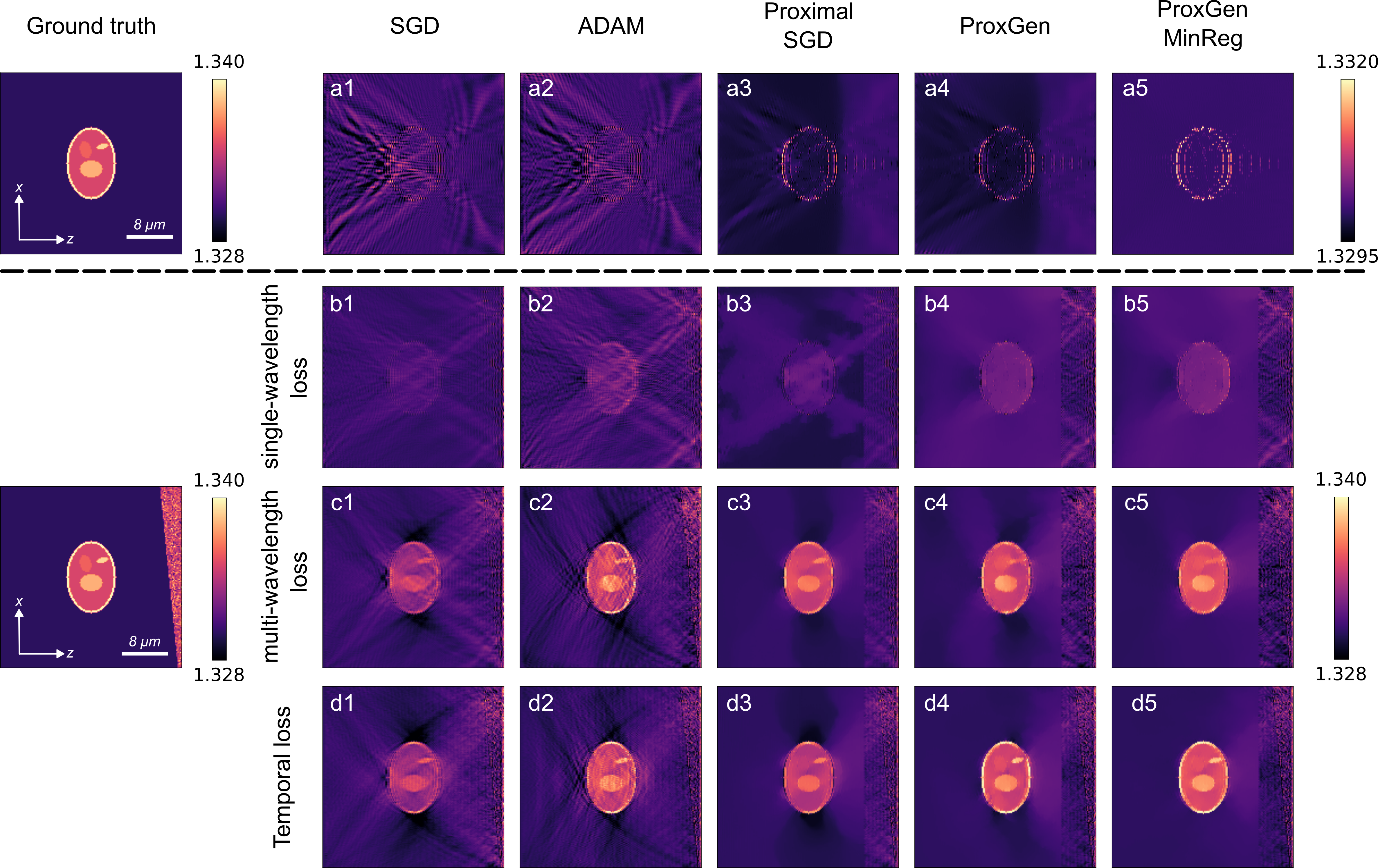}
    \caption{2D reconstruction results for different configurations with order 5 Born forward model. Each column corresponds to an optimization algorithm. Row (a) shows the results for the single-wavelength loss without a reflector. Since the results for different loss functions without reflector are similar, they are not displayed here. The last three rows present the reconstructions using different loss functions with ground truth including a reflector. The unweighted temporal loss is not shown, as it produces results similar to the multi-wavelength loss (row (c)): the Fourier transform does not significantly impact the results. The weighted temporal loss shown here corresponds to equation \eqref{temporal loss} with a coefficient $\gamma = 0.1$.}
    
    \label{fig:fig3}
\end{figure*}

Additionally, a differentiable regularization term, referred to as \textit{Minimum regularization}, is introduced. This term prevents predictions below the RI $n_{\text{b}}$, acting as a Ridge regression for points with a RI lower than $n_{\text{b}}$. Since this regularization is differentiable, it can be easily integrated into the temporal loss function. This additional regularization corresponds to a positivity constraint. Choi \textit{et al.} \cite{choi2007tomographic} actually demonstrate that it reduces the underestimation of the RI in a transmission configuration. 
For the same reason as the TV regularization, minimum regularization is then only applied to the $V_1$-object region (contained in the layer $[z_1,z_2]$ in Fig.~\ref{fig_1}(a)) but not to the $V_2$-reflector (layer $[z_2,z_3]$ in {Fig.~\ref{fig_1}(a)).

Figure \ref{fig:fig3} illustrates the improvement in the object reconstruction achieved by applying the aforementioned strategies. The first row (Fig. \ref{fig:fig3}(a)) highlights the challenge of reconstructing the sample without a reflector, even when using various iterative optimization techniques. In this case, the scattering potential remains trapped in a local minimum or flat region, preventing accurate reconstruction. As previously highlighted, the low-spatial frequencies of the object can be hardly retrieved in such a configuration. 
However, the second row (Fig. \ref{fig:fig3}(b)) shows that simply adding a reflector is not sufficient if the temporal aspect is not considered. In a single-wavelength experimental setup, {the singly-scattered term} $U_1$ and $U_2$, as well as the second-order crossed terms $U_{12}$ and $U_{21}$, cannot be disentangled. This mixing prevents the convergence of the optimization algorithms.
The last two rows of Fig. \ref{fig:fig3} demonstrate that temporal resolution is essential for accurate cell reconstruction. The multi-wavelength loss (Fig. \ref{fig:fig3}(c)) corresponds to a broadband loss function that incorporates multiple wavelengths: 

\begin{align}
\label{parseval1} 
\mathcal{L}_{B}(\mathbf{V}) =  \sum_{\omega} \| \mathbf{y}_{\text{Born-}i}(\mathbf{V}, \mathbf{E}_{\text{in}}, \omega ) -\mathbf{E}_{\text{m}}(\omega) \|^2.
\end{align}

Due to Parseval's theorem, $\mathcal{L}_{B}(\mathbf{V})$ is actually equivalent to a temporal loss integrated over all times of flight: 

\begin{align}
\label{parseval2} 
\mathcal{L}_{B}(\mathbf{V}) =  \sum_{t} \| \mathbf{y}_{\text{Born-}i}(\mathbf{V}, \mathbf{E}_{\text{in}}, t ) -\mathbf{E}_{\text{m}}(t) \|^2 
\end{align}

All time points are here weighted equally. While this approach enables good cell reconstruction, further improvement can be achieved using a weighted temporal loss, as described in \eqref{reflection loss} and illustrated by Fig.~\ref{fig:fig3}(d). 
Thus, employing a multi-wavelength configuration and measuring the amplitude and phase of the reflected wavefield ( and not just the intensity) are critical to reducing the ill-posedness of the inverse problem. Indeed, it gives access to the time-dependence of each reflected wave-field  which greatly facilitate the reconstruction.

\begin{figure}[htbp]
    \centering
    \includegraphics[width=1.0\linewidth]{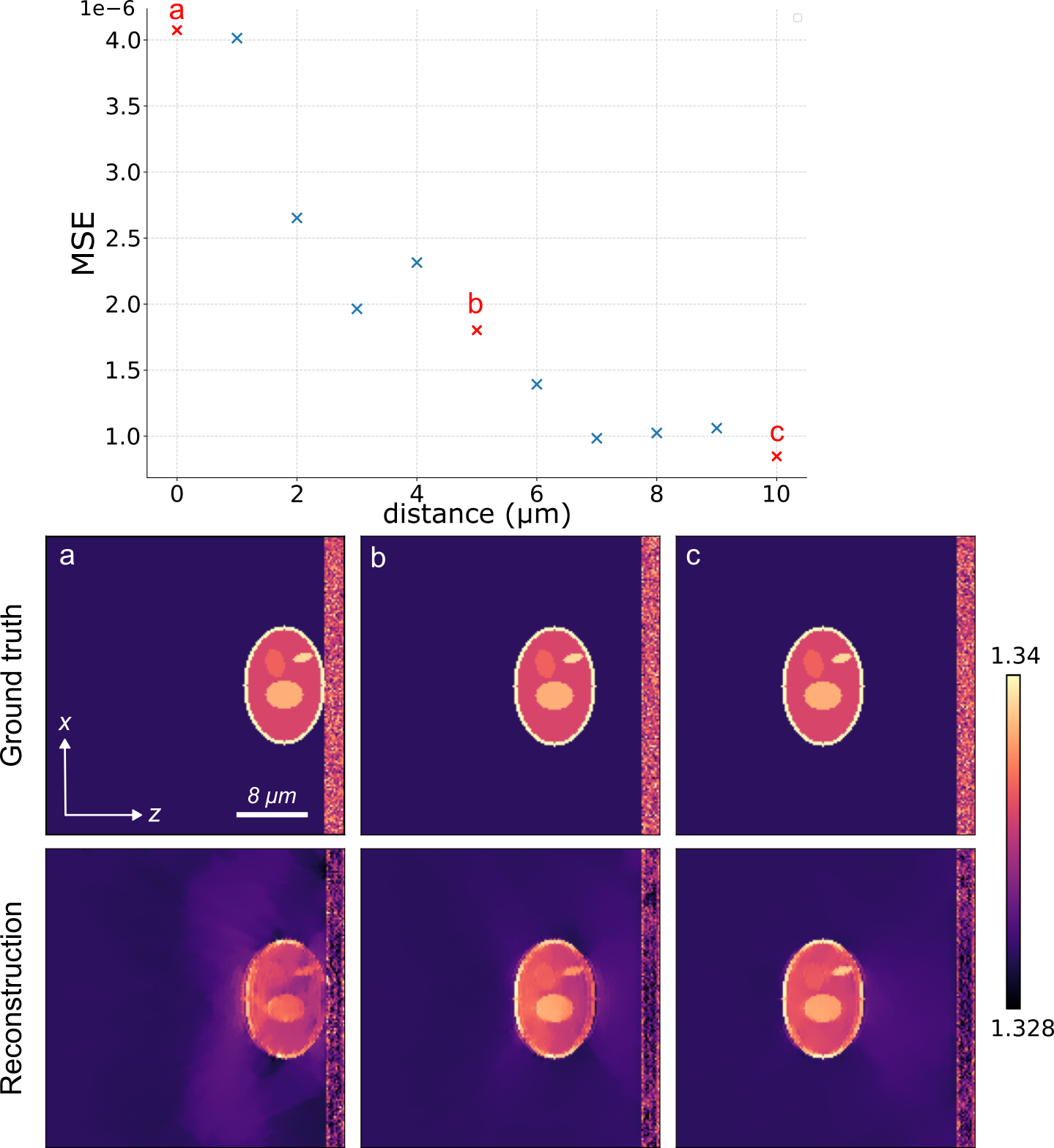}
    \caption{Effect of the distance between reflector and cell on reconstruction quality. The mean squared error (MSE), calculated only inside the cell, shows how the reconstruction differs from the ground-truth cell as a function of distance. Reconstructions at different distances are shown to illustrate how reconstruction quality decreases as the cell and reflector get closer.}
    \label{fig:fig9}
\end{figure}

The TV regularization is here essential to reconstruct correctly the sample as the reconstruction is noisy without it. While SGD and Adam-based gradient descent methods yield similar results for the multi-wavelength loss ((Figs. \ref{fig:fig3}(c3) and \ref{fig:fig3}(c4)), Adam-based methods perform better when using the weighted temporal loss ((Figs. \ref{fig:fig3}(d3) and \ref{fig:fig3}(d4)). The results improve further when applying minimum regularization, which prevents the RI from falling below $n_{\text{b}}$.

In our reconstruction framework, the foreground object and the background reflector are spatially separated for two main reasons. First, our theoretical model assumes that evanescent waves from the reflector are negligible. Second, the temporal separation helps the optimization algorithm converge more reliably as mentioned above. However, the object can still be reconstructed (even if less accurately) when the spatial separation is limited, as described in Fig.~\ref{fig:fig9}.
As expected, the reconstruction quality decreases as the reflector gets closer, dropping below 1µm. Since the measured signal is included in $[800nm,875nm]$, the temporal resolution of the reconstructed wave packet is $d_{z,\mathrm{temp}} \approx 3.5 \mu\mathrm{m}$. At distances around $1\mu m$, the cross-interaction and direct reflection terms from Eq.~\eqref{U second born} begin to overlap, causing mixing that limits reconstruction performance.

\subsubsection{Importance of a weighted temporal loss}
\label{subsecimportanceweightedloss}%
Unlike alternative methods that reconstruct the RI of an object using a deterministic reflector \cite{li2024reflection}, our approach does not require any prior knowledge on it. Indeed, this reflector can belong to the inspected medium. The RI distribution can be reconstructed, even when the background reflector is highly irregular and initially unknown. 

\begin{figure}[htbp]
    \centering
    \includegraphics[width=1.0\linewidth]{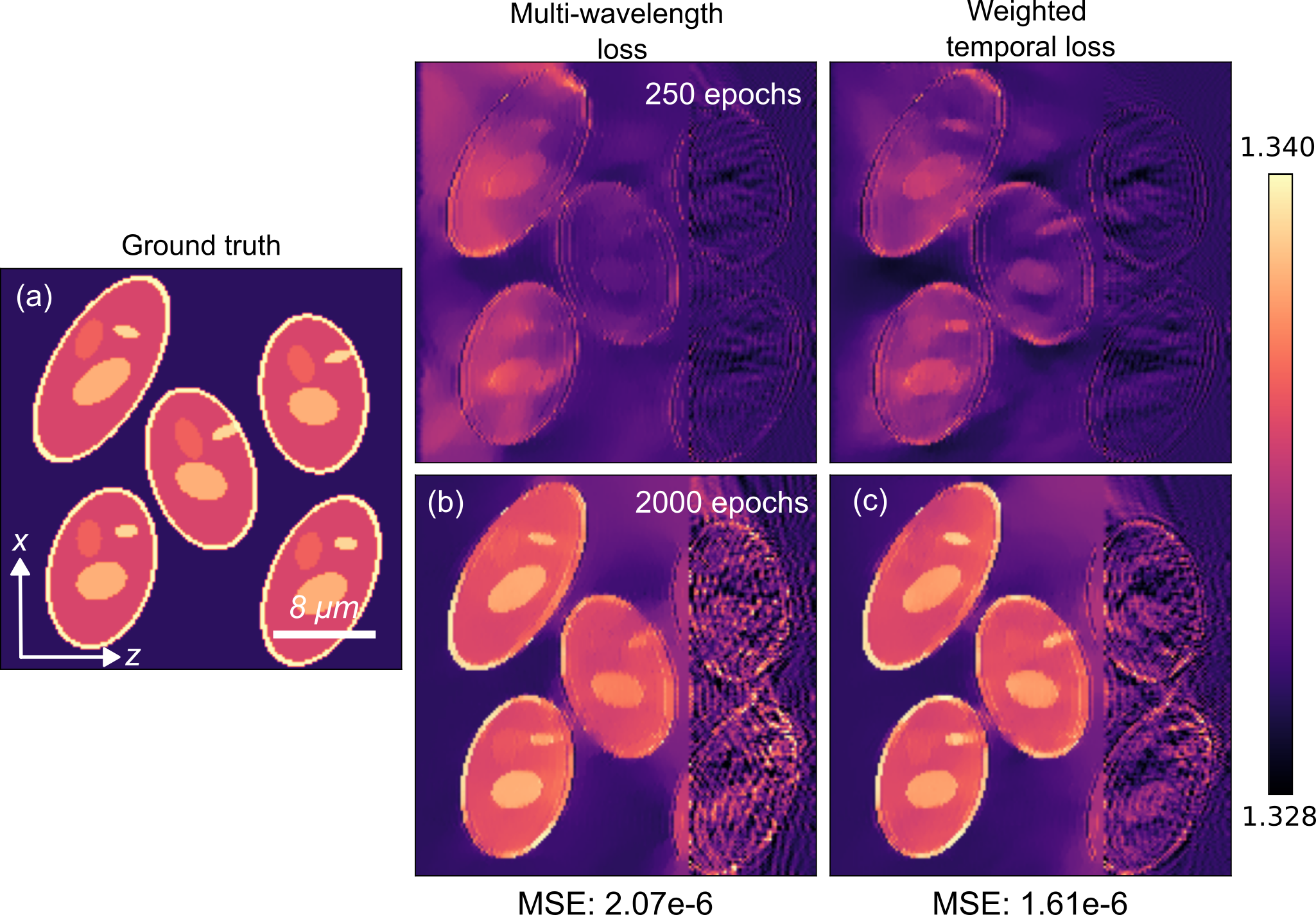}
    \caption{Reconstruction of complex samples thanks to the \textit{ProxGen algorithm} by using different loss functions. Here the reflector corresponds to the background cells at the right. The regularization (TV by proximal and minimum regularization) are applied at the left of the image and not for the reflector. As in previous cells the initialization corresponds to uniform medium. The reconstruction is shown both at an early stage of training (250 epochs) and at the final stage (2000 epochs), highlighting that the temporal loss improves reconstruction stability and consequently reduces the dependence on regularization. The MSE represents the mean squared error computed inside the cells at the end of training.}
    \label{fig:fig5}
\end{figure}

Additionally, this algorithm can be applied to more complex samples. For instance, we can consider a 2D sample made up of multiple cells, and try to reconstruct the upstream cells by using the background cells as reflectors.
Examples of such reconstructions are shown in Fig.~\ref{fig:fig5}, where we can observe that the weighted temporal loss (Fig.~\ref{fig:fig5}(c)) leads to more satisfactory results and shows noticeably better reconstruction even at early stages of convergence compared to the multi-wavelength loss (Fig.~\ref{fig:fig5}(b)) [see the comparison with ground truth displayed in Fig.~\ref{fig:fig5}(a)]. For both losses (multi-illumination and weighted temporal), the low and high spatial frequency components are effectively separated in time, thanks to Parseval's theorem. However, the multi-illumination loss treats all times equally, because the energy received at both echo times is similar, the model struggles to reconstruct the low-frequency components of the object, which are mainly carried by the weaker $U_{12}$ and $U_{21}$ terms rather than by the dominant $U_1$ term. 
On the contrary, the weighted temporal loss enhances the strength of backscattered echoes and leads to a better image the sample.

Thus, despite being far from the ideal case of a planar reflector, our method enables high-quality reconstruction of RI maps. Moreover, the use of a weighted loss allows for the reconstruction of both low and high spatial frequencies, effectively mitigating the missing cone problem commonly encountered in transmission. However, as shown in Fig.~\ref{fig:fig3}, achieving a good reconstruction requires the presence of a reflector (and therefore a complex medium), as well as a temporal loss (hence a multi-illumination device), unlike in a transmission configuration.

\subsubsection{3D reconstruction} 

\begin{figure*}[htbp]
    \centering
    \includegraphics[width=1.0\linewidth]{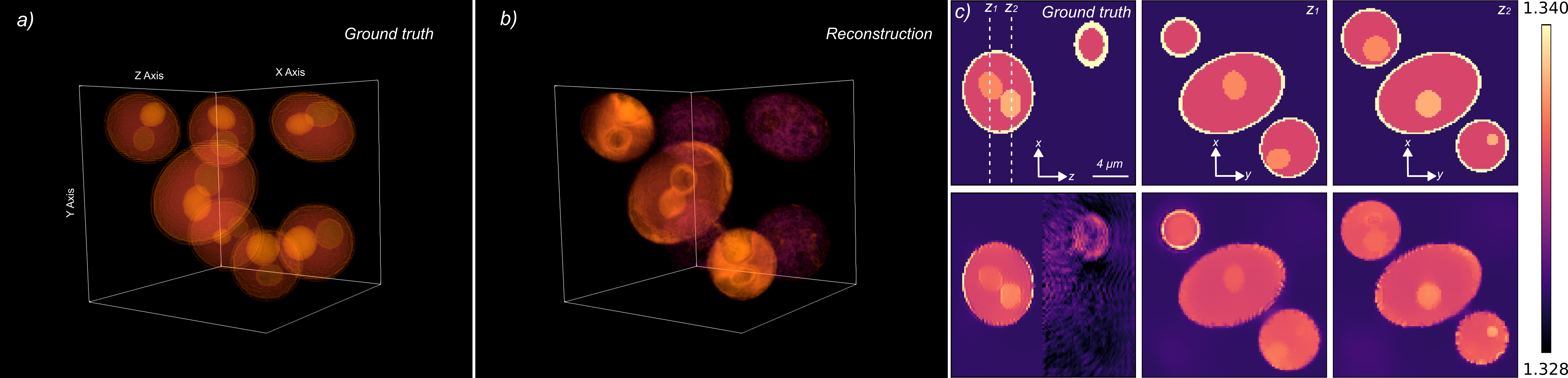}
    \caption{3D Reconstruction with the proposed method of $20 \times 20 \times 20 \, \mu m$ sample made of several cells. Figure (a) shows the ground truth 3D rendering of the RI map, figure (b) corresponds to the reconstruction and figure (c) shows a series of ground truth and prediction 2D cuts. The inhomogeneity of the reconstructed cells in figure (b) is due to the fact that the reconstructed background cells correspond to an equivalent reflectivity of lower optical index than the ground truth. An animation is available in the supplementary material.}
    \label{fig:fig7}
\end{figure*}
To demonstrate the scalability of our model, we also simulate 3D reconstructions. The sample under study is displayed in  Fig.~\ref{fig:fig7}(a). The simulation domain is a cube of size $N_x = N_y = N_z = 101$ of resolution $dx=dy=dz=200$ nm. The simulated illumination sequence involves 80 illuminations spirally distributed within a pupil of numerical aperture $\text{NA} = 0.8$. Each illumination is repeated for 11 wavelengths uniformly distributed between 800 and 875 nm.  
The forward model used in our optimization process is still a 5$^{\textrm{th}}$ order Born model since we remain in a weakly multiple scattering regime. The optimization is performed over 1000 epochs to ensure convergence, although a good reconstruction can already be achieved after 400 epochs. The reconstruction was efficiently carried out on an RTX 6000 Ada GPU with 48GB of VRAM. Each epoch took 30 seconds, which is longer than for an equivalent transmission reconstruction, as wave propagation needs to be computed for multiple wavelengths. 

Due to the long training time, we fine-tuned our optimization algorithm on a smaller cube and applied the same hyper-parameters to the larger one. We used a learning rate of $\alpha = 2 \times 10^{-5}$, momentum parameters $(\beta_1, \beta_2) = (0.9, 0.999)$, and a weight factor $\gamma = 0.1$ to generate the figure. The result of the optimization process is displayed in Fig.~\ref{fig:fig7}.
It appears that in 3D, it is still possible to obtain good reconstructions of cells in the foreground, by using cells in the background. The average values of the various substructures remain consistent with the ground truth. However, the reconstruction quality is not as good as in 2D, due to the larger number of parameters to optimize and the increased difficulty in tuning the hyperparameters.

\section{Conclusion}

We have introduced a novel optimization algorithm for 3D refractive index reconstruction in reflection-mode microscopy. Our method is based on the exploitation of multiple light scattering.  More precisely, our approach consists in leveraging the light backscattered by deep reflectors to map the refractive index distribution in the foreground. The background reflector is \textit{a priori} unknown, and its reflectivity is simultaneously reconstructed as part of the training process. This approach is therefore a promising candidate to reach quantitative in vivo optical imaging.

The ill-posed character of the problem is circumvented by a weighted temporal loss in order to focus on the low frequency part of the object and by using positivity constraint and sparsity regularization to converge to a good minimum. The resulting reconstructions closely match the ground truth across both low and high spatial frequencies, highlighting the effectiveness of the weighted temporal loss.

We implemented a 2D and 3D version to demonstrate its effectiveness in simulations. To remain consistent with the theory, we restricted our study to weakly scattering media and used relatively simple forward models compared to transmission ODT \cite{kamilov2016recursive}. However, in practice, this method could involve more advanced forward models for larger refractive index contrast as Multi-layer Born \cite{chen2020multi}, modified Born series \cite{modifiedBorn} or Lippmann-Schwinger Model \cite{pham2020three}. Whether it can be adapted to a more strongly scattering regime still needs to be demonstrated, either in simulation or experimentally.

We have also demonstrated the necessity of multi-wavelength analysis in reflection to achieve the required temporal approach. As a result, ODT in reflection will be more time-expensive than transmission approaches but will be rewarding since we will have access to both the low and high spatial frequencies of the sample. This also helps overcome the missing cone issue inherent to transmission configurations.

Beyond optical microscopy, the proposed approach can be leveraged to map the local wave speed in any field of wave physics for which a frequency- or time-resolved measurement of the reflection matrix is possible~\cite{bureau_three-dimensional_2023,giraudat_unveiling_2023}. Indeed, not only is the local wave speed a quantitative marker for ultrasound diagnosis~\cite{Staehli2023} but it is also, for instance, an important monitoring parameter of tectonic motion~\cite{Solarino2024} and volcanic eruption~\cite{DAuria2022} in geophysics. In each of this field, multiple scattering is far from being negligible~\cite{Chaput2015,Goicoechea2024} and its inclusion in the forward model would be extremely rewarding for quantitative imaging. This is the principle of full wave-form inversion~\cite{Virieux2009} but we here show how such brute force strategies would benefit from physically-grounded loss functions and simplifications of the forward model in order to guarantee the convergence of the inversion process.

\section*{Acknowledgments}
A.A. is grateful for the funding provided by the European Research Council (ERC) under the European Union's Horizon 2020 research and innovation program (grant agreement no. 819261, REMINISCENCE project). 
The work of J.G. was supported in part  by the Agence de l'Innovation de D\'efense (AID) via Centre Interdisciplinaire d'\'Etudes pour la D\'efense et la S\'ecurit\'e (CIEDS) project PRODIPO.

\appendix
\begin{center}
\textbf{Appendix A}\\[0.5em]
\textbf{Filtering operation on source term}
\end{center}
\label{appendix:A}

This section is a proof of the filtering effect of measurement in transmission and reflection configuration. The proof is the same as in \textit{Emil Wolf} \cite{firstborn} work, but without the first order Born hypothesis.\\

We express the scattered field as a function of the source term:
{\small
\begin{align}
E_{\text{s}}(&x,y,z) =  \iiint_{\mathbb{R}^3} U(x', y', z') \nonumber \notag \\
&\quad \times G(x-x', y-y', z-z') \, dx'dy'dz' \tag{\ref{convgreen}}
\end{align}
}

Since the object to be imaged is of finite size, we assume that $V$ is supported in $[z_1,z_3]$ and that we measure the scattered field for $z \in (-\infty,z_1) \cup (z_3,+\infty) $. \\
Since we are measuring $E_{\text{s}}$ at a fixed $z$, it is natural to look for calculations based on the 2D Fourier transform of the 3D Green's function. Thus we have \cite{weyl1919ausbreitung}:

{\small
\begin{align}
\label{greenft2d}
  \iint_{\mathbb{R}^2}
G(x,y,z) e^{-j ( k_x x + k_y y)} dx dy  
= \frac{j}{2\kappa(k_x,k_y)}e^{j \kappa(k_x,k_y)|z|} \notag \\
G(x,y,z) = \iint_{\mathbb{R}^2}  \frac{j}{2\kappa( k_x,k_y)}e^{j(k_xx+k_yy+ \kappa(k_x,k_y)|z|)} dk_xdk_y
\end{align}
}

with $\kappa$ defined as:

{\small
\begin{align*}
\kappa(k_x, k_y) = 
\begin{cases} 
\sqrt{n_{\text{b}}^2k_{0}^{2} - k_{x}^{2} - k_{y}^{2}} & \text{if } n_{\text{b}}^2k_0^2 \geq k_x^2 + k_y^2, \\
j\sqrt{k_{x}^{2} + k_{y}^{2} - n_{\text{b}}^2k_{0}^{2}} & \text{if } n_{\text{b}}^2k_0^2 < k_x^2 + k_y^2.
\end{cases}
\end{align*}
}
$\kappa$ is the $z$-wavelength of the Green function with two modes: a propagative mode and an evanescent one.\\
In the following, we rewrite \eqref{greenft2d} taking into account the fact that $z \in (-\infty,z_1) \cup (z_3,+\infty) $ and $z' \in \text{supp}(V)$
{\small
\begin{align}
\label{eq: G tf 2d relative}
G(&x-x',y-y',z-z') = \iint_{\mathbb{R}^2}  \frac{j}{2\kappa( k_x,k_y)} \notag \\ 
&\quad \times e^{j(k_x(x-x')+k_y(y-y')+ \text{sgn}(z)\kappa(k_x,k_y)(z-z'))} dk_xdk_y
\end{align}
}
By injecting \eqref{eq: G tf 2d relative} into \eqref{convgreen} we obtain:
{\small
\begin{align} \label{Es and A}
E_{\text{s}}(x,y,&z) = \iint_{\mathbb{R}^2} A(k_x, k_y, z) \notag \\
& \times e^{j(k_x x +k_y y+\text{sgn}(z)\kappa(k_x,k_y)z)} \, dk_xdk_y\\
\label{Es and A 2}
A(k_x, k_y, &z) = \frac{j}{2\kappa(k_x, k_y)}\iiint_{\mathbb{R}^3}  U(x', y', z') \notag \\
& \times e^{-j(k_xx'+k_yy' + \text{sgn}(z)\kappa(k_x,k_y)z')} dx'dy'dz'
\end{align}
}

By developing \eqref{Es and A}, we can express $A$ by the 2D Fourier transform of $E_{\text{s}}(.,z)$
{\small
\begin{align} \label{Es and A 3}
\mathcal{F}_{2D}(E_{\text{s}}(\cdot,z))(k_{x}, k_{y})e^{-\text{sgn}(z)j\kappa(k_x, k_y)z} = A(k_x, k_y, z)
\end{align}
}
The equation is difficult to express in simple terms, because $\kappa$ is a complex number. However, we assume that we are measuring with $z$ far enough away from the support of $V$ that the evanescent waves are negligible (i.e. $A(k_x, k_y, k_z) \approx 0$ if $n_{\text{b}}^2 k_0^2 < k_x^2 + k_y^2$). Consequently, $\kappa$ is real, and we can express $A$ directly as the Fourier transform of $U$:

{\small
\begin{align} \label{A and U}
A(k_x, k_y, z) = \frac{j\tilde{U}(k_x, k_y, \text{sgn}(z)\sqrt{n_{\text{b}}^2k_0^2-k_x^2-k_y^2})}{2\sqrt{n_{\text{b}}^2k_0^2-k_x^2-k_y^2}}
\end{align}
}

And by combining \eqref{A and U} and \eqref{Es and A 3} we obtain:

{\small
\begin{align}
\mathcal{F}_{2D}(E_{\text{s}}(\cdot,z))(k_{x}, k_{y}) \propto \tilde{U}(k_x, k_y, \mathrm{sgn}(z)\sqrt{n_{\text{b}}^2k_0^2-k_x^2-k_y^2}) \tag{\ref{Ewald sphere}}
\end{align}
}
We finally found that we can only measure the Ewald sphere of the source $U$ as shown in \cite{coupland2008holography} under the Fraunhofer approximation.

\begin{center}
\textbf{Appendix B}\\[0.5em]
\textbf{Filtering operation under second Born approximation}
\end{center}
\label{appendix:B}

Let us begin by examining the $U_{12}$ term, which represents an interaction with the reflector $V_2$ followed by an interaction with the potential $V_1$. We can express $U_{12}$ as:

{\small
\begin{align} \label{U12 with H2}
U_{12}(x,y,z) &= V_1(x,y,z)H_2(x,y,z) \\
H_{2}(x,y,z) &= \iiint_{\mathbb{R}^3} G(x-x',y-y', z-z') \notag \\ 
& \times V_{2}(x', y', z') E_{\text{in}}(x',y',z')\, dx'dy'dz'  \notag \\
&= \iint_{\mathbb{R}^2} G(x-x',y-y', z) \notag \\ 
& \times v_{2}(x', y') E_{\text{in}}(x',y',0)\, dx'dy' \notag
\end{align}
}
$H_2$ corresponds to a new incident wave from the reflector $V_2$. \\
By injecting the 2D Fourier transform of the Green function \cite{weyl1919ausbreitung} \eqref{greenft2d}, assuming that $z$ is small enough to neglect evanescent waves (since $\text{supp}(V_1) = [z_1, z_2]$ we assume here that $z_2$ is far enough from 0) and by using convolution theorem, we can express $H_2$ in a 2D integral form:

{\small
\begin{align}
\mathcal{F}_{2D}(H_2(.,z)) &= 
\frac{j e^{j\sqrt{n_{\text{b}}^2 k_0^2 - k_x^2 - k_y^2} |z|}}%
{2\sqrt{n_{\text{b}}^2 k_0^2 - k_x^2 - k_y^2}} \,
\tilde{v_2}(k_x - k_{\text{in},x}, k_y - k_{\text{in},y}) \nonumber \\
H_2(x,y,z) &= \iint_{\mathbb{R}^2}
\frac{j \, \tilde{v_2}(k_x - k_{\text{in},x}, k_y - k_{\text{in},y})}%
{2\sqrt{n_{\text{b}}^2 k_0^2 - k_x^2 - k_y^2}} \,
e^{j(k_x x + k_y y)} \nonumber \\
&\quad \times
e^{j \sqrt{n_{\text{b}}^2 k_0^2 - k_x^2 - k_y^2} |z|} \, dk_x \, dk_y \label{F2H2}
\end{align}
}

By applying a Fourier transform to \eqref{U12 with H2}, substituting \eqref{F2H2}, and simplifying the absolute value due to the support of \( V_{1}(., z') \) being restricted to \( (-\infty,0) \), we obtain:

{\small
\begin{align}
&\tilde{U}_{12}(k_x,k_y,k_z) = \iint_{\mathbb{R}^2} \frac{\tilde{j v_{2}}(k'_x-k_{\text{in},x}, k'_y-k_{\text{in},y})}{2\sqrt{n_{\text{b}}^2k_{0}^{2} - k'_x{}^{2} - k'_y{}^2}} \tag{\ref{U12}} \\
&\times \tilde{V_1}(k_x - k'_x, k_y - k'_y, k_z + \sqrt{n_{\text{b}}^2k_0^2-k'_x{}^2 - k'_y{}^2})\, dk'_x dk'_y \notag
\end{align}
}

The second term $U_{21}$ corresponds to the interaction with $V_1$ followed by the interaction with the reflector $V_2$.

{\small
\begin{align} \label{U21 with H1}
U_{21}(x,y,z) &= V_2(x,y,z)H_1(x,y,z) \\
H_1(x,y,z) &= \iiint_{\mathbb{R}^3} G(x-x',y-y', z-z') \notag \\ 
& \times V_{1}(x', y', z') E_{\text{in}}(x',y',z')\, dx'dy'dz' \notag
\end{align}
}
By applying a Fourier transform to $U_{21}$ we have:

{\small
\begin{align*}
\tilde{U}_{21}(k_x,k_y,k_z) &= \iiint_{\mathbb{R}^3}  V_2(x,y,z)H_1(x,y,z) \\
&\times e^{j(k_xx+k_yy+k_zz)}dxdydz \\
&= \iint_{\mathbb{R}^2}  v_2(x,y)H_1(x,y,0)e^{j(k_xx+k_yy)}dxdy \\ 
&= \iint_{\mathbb{R}^2} \tilde{v_2}(k_x-k_x',k_y-k_y') \\
&\times \mathcal{F}_{2D}(H_1(.,z=0))(k_x',k_y')dk_x'dk_y'
\end{align*}
}
In order to get an analytical expression for the Fourier transform $U_{12}$, we need to express the 2D Fourier transform of $H_1$:

{\small
\begin{align*}
&\mathcal{F}_{2D}(H_{1}(., z=0))(k_x, k_y) \\
&=
\iint_{\mathbb{R}^2} H_{1}(x,y,0) e^{- j( k_x x + k_y y)} dx dy
\\
&= 
\iiint_{\mathbb{R}^3} \iint_{\mathbb{R}^2}
G(x-x',y-y',-z') 
 V_1(x',y',z') \\
&\times e^{j(k_{\text{in},x}x' +k_{\text{in},y}y' +k_{\text{in},z}z') -j ( k_x x + k_y y)}dx dy dx'dy'dz'   \\
&= 
\iiint_{\mathbb{R}^3}  \iint_{\mathbb{R}^2}
G(x'',y'',-z') e^{-j ( k_x x'' + k_y y'')} dx'' dy'' 
 \\
&\times  V_1(x',y',z') e^{j[(k_{\text{in},x}-k_x )x' +(k_{\text{in},y}-k_y) y' +k_{\text{in},z}z') }dx'dy'dz' 
\\
&= 
\int_{\mathbb{R}} dz' \Big[  \iint_{\mathbb{R}^2}
G(x'',y'',-z') e^{-j ( k_x x'' + k_y y'')} dx'' dy''  \Big]
 \\
&\times \Big[ \iint_{\mathbb{R}^2}  V_1(x',y',z') e^{j[(k_{\text{in},x}-k_x )x' +(k_{\text{in},y}-k_y) y' +k_{\text{in},z}z'] }dx'dy'\Big]
  \end{align*}
}
The first term between square brackets is the 2D Fourier transform of the Green function and defined in eq \eqref{greenft2d}. The second term between square brackets is:

{\small
\begin{align*}
\iint_{\mathbb{R}^2}  V_1(x',y',z') e^{j [(k_{\text{in},x}-k_x )x' +(k_{\text{in},y}-k_y) y' +k_{\text{in},z}z'] } dx'dy' \\
=
\mathcal{F}_{2D}(V_{1}(., z')) (k_x-k_{\text{in},x}, k_y-k_{\text{in},y})  
e^{j k_{\text{in},z} z'}
\end{align*}
}

Since the support of $V_{1}(., z')$ is in $(-\infty,0)$, the integral in $z'$ is concentrated on $(-\infty,0)$ and:

{\small
\begin{align*}
&\mathcal{F}_{2D}(H_{1}(., z=0))(k_x, k_y) = \\
&\int_{\mathbb{R}}
 \frac{j\mathcal{F}_{2D}(V_{1}(., z')) (k_x-k_{\text{in},x}, k_y-k_{\text{in},y})  }{2 \kappa(k_x, k_y)}e^{j (k_{\text{in},z} -\kappa(k_x, k_y)) z'} dz'
\end{align*}
}
which gives (by ignoring evanescent waves):
{\small
\begin{align} 
\label{FT 2D H_1}
&
\mathcal{F}_{2D} (H_{1}(., z=0)) (k_x, k_y)
=
 \frac{j}{2 \sqrt{n_{\text{b}}^2k_{0}^2-k_x^2-k_y^2}} \notag
 \\
 &  \times \tilde{V_{1}} (k_x-k_{\text{in},x}, k_y-k_{\text{in},y} , \sqrt{n_{\text{b}}^2 k_{0}^2-k_x^2-k_y^2 } - k_{\text{in},z} )
\end{align}
}
Finally, we obtain the following formula for $U_{21}$:
{\small
\begin{align}
&\tilde{U}_{21}(k_x,k_y,k_z)= \iint_{\mathbb{R}^2} \frac{\tilde{j v_{2}}(k_x-k'_x, k_y-k'_y)}{2\sqrt{n_{\text{b}}^2k_{0}^{2} - k'_x{}^{2} - k'_y{}^2}} \tag{\ref{U21}}\\
& \times \tilde{V_1}(\tilde{k}_x-k_{\text{in},x}, \tilde{k}_y-k_{\text{in},y}, \sqrt{n_{\text{b}}^2k_0^2-k_x^2 - k_y^2}-k_{\text{in},z})\, dk'_x dk'_y \notag
\end{align}

\begin{center}
\textbf{Appendix C}\\[0.5em]
\textbf{Noise analysis, selection of hyperparameters and regularization}
\end{center}
\label{appendix:C}

Since we had to solve an ill-posed inverse problem, longer training times were necessary. To ensure better convergence for method comparison, we fine-tuned the learning rate and trained until 
2000 epochs for Figs. \ref{fig:fig4}-\ref{fig:fig5}. However, the reconstruction quality becomes acceptable around 500 epochs, even though the loss continues to decrease. An illustration of the convergence is presented Fig.\ref{fig:fig12}.

\begin{figure}[htbp]
    \centering
    \includegraphics[width=1.0\linewidth]{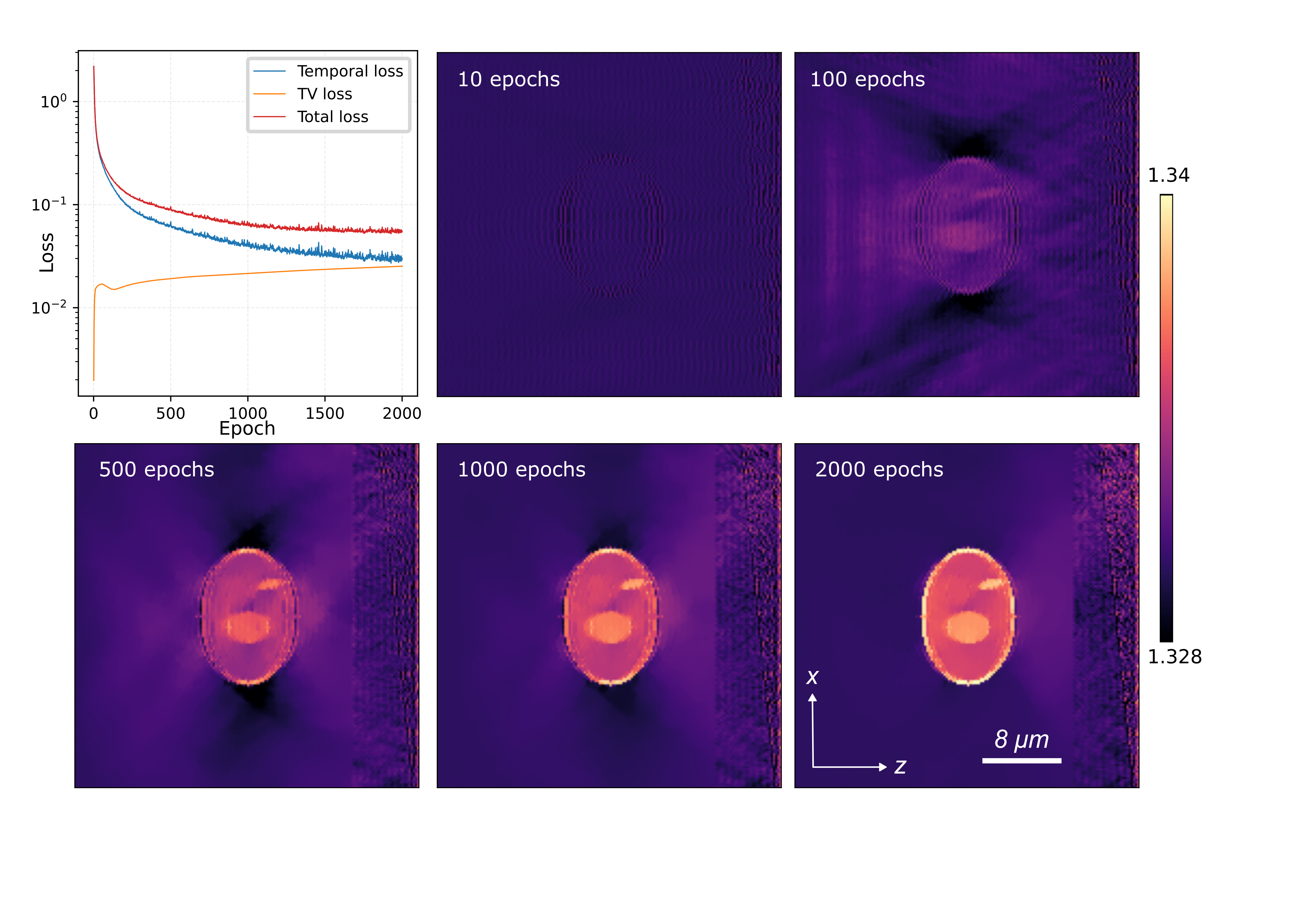}
    \caption{ Analysis of the convergence process. During the first epochs, only the high-frequency components are reconstructed, while the low-frequency information of the object gradually appears around the 500th epoch. The TV regularization parameter is small to allow the recovery of as much information as possible before the regularization term becomes dominant, leading to a better reconstruction.}
    \label{fig:fig12}
\end{figure}

The optimal learning rate was generally found to be around $\alpha = 10^{-5}$. For adaptive methods, the coefficients were set as recommended in \textit{Kingma's et al} work \cite{kingma2014adam} $(\beta_1, \beta_2) = (0.9, 0.999)$. The weight factor was set to $\gamma = 0.1$ for all simulations and kept unchanged across different cases, as this ensured consistently good performance, but could be fine tuned depending on the medium to be reconstructed. The principle is that one tries to balance the contributions of the direct reflection from the foreground object (that contains the high spatial frequencies on the foreground object) and the cross-interaction terms (that contains the low spatial frequencies on the foreground object). Decreasing $\gamma$ reduces the weight of the direct-reflection component, which corresponds to information that is reconstructed rapidly during optimization (see Fig.\ref{fig:fig12}), thereby helping the optimization focus on the more challenging cross-interaction terms. Keeping $\gamma > 0$ is generally preferable, since the high-frequency information helps mitigate the missing-cone problem by adding complementary data. TV regularization and minimum regularization are not applied on the reflector because it allows better reconstruction of foreground objects as shown in Fig.\ref{fig:fig11}. \\

\begin{figure}[htbp]
    \centering
    \includegraphics[width=1.0\linewidth]{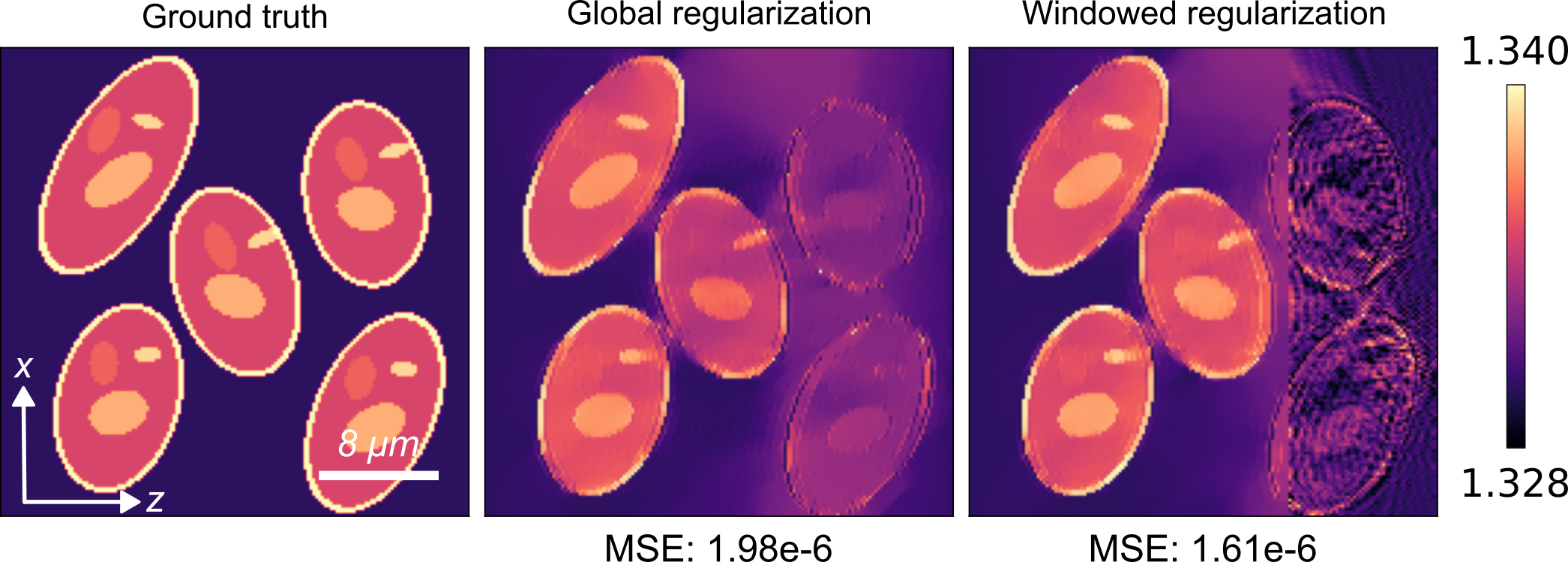}
    \caption{Reconstruction of a multi-cell medium under different regularization strategies. The comparison is made between a reconstruction with regularization applied to the entire medium and  applied only to the foreground region. The Mean Squared Error (MSE) of the reconstruction inside the foreground cells is displayed.}
    \label{fig:fig11}
\end{figure}

The reconstruction quality is slightly better when no regularization is applied to the reflector, as this avoids constraining on it. In this case, the reconstructed RI of the reflector corresponds to an effective value that produces better reflectivity, rather than to its true physical index. Because the reflector’s low-frequency components are difficult to retrieve, this effective value mainly acts as a local minimum that mimics the reflective behavior.
Adding regularization, in this case, limits the reflector without bringing useful physical information about it.\\

\begin{figure}[htbp]
    \centering
    \includegraphics[width=0.9\linewidth]{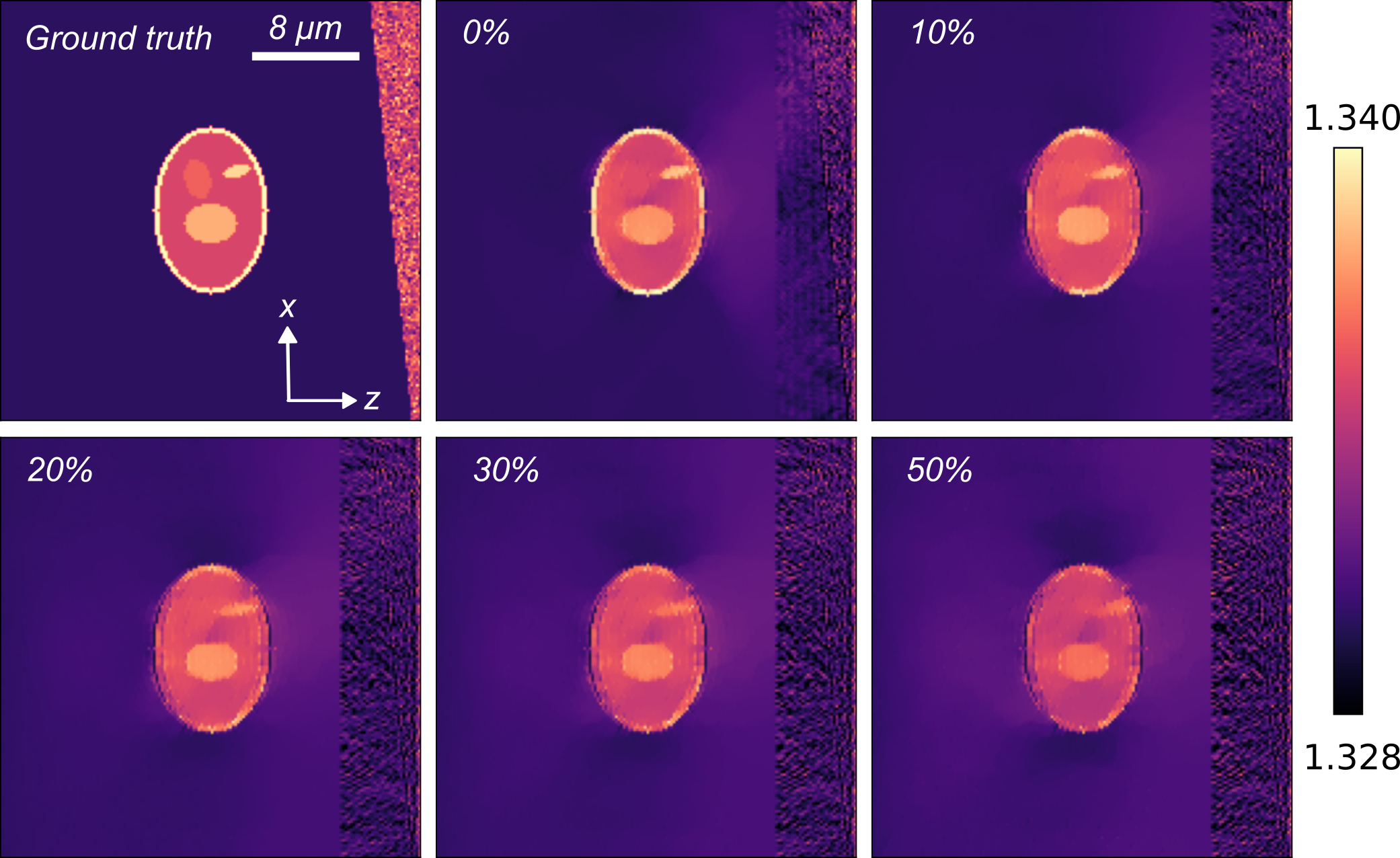}
    \caption{Evolution of reconstruction as a function of the level of Gaussian noise added to the simulated measurements.  Despite the loss of precision, the 2D reconstruction of the foreground cell remains consistent with the ground truth. Reconstructions are performed with weighted temporal loss ($\gamma =0.1$) with \textit{ProxGen} algorithm and minimum regularization.}
    \label{fig:fig6}
\end{figure}

Despite the use of the weighted temporal loss to focus on the low-frequency signal, the optimization problem is not straightforward and requires more training time than in transmission. Thus our optimization algorithm could be sensitive to noisy measurement. To verify this, we conducted 2D tests by adding noise to the ground truth measurements. As shown in Figure \ref{fig:fig6}, our model remains robust to additive Gaussian noise up to $50 \%$. The noise level is here defined relatively to the mean of the reflected absolute amplitude.

\begin{center}
\textbf{Appendix D}\\[0.5em]
\textbf{Spectral study of the reconstruction}
\end{center}
\label{appendix:D}

In order to validate the importance of cross-terms $U_{12}$ and $U_{12}$ described in Eq.~\eqref{U21} and Eq.~\eqref{U12}, we display in Fig.~\ref{fig:fig10} the spectral representation of the reconstructed potential $V$.

\begin{figure}[htbp]
    \centering
    \includegraphics[width=1.0\linewidth]{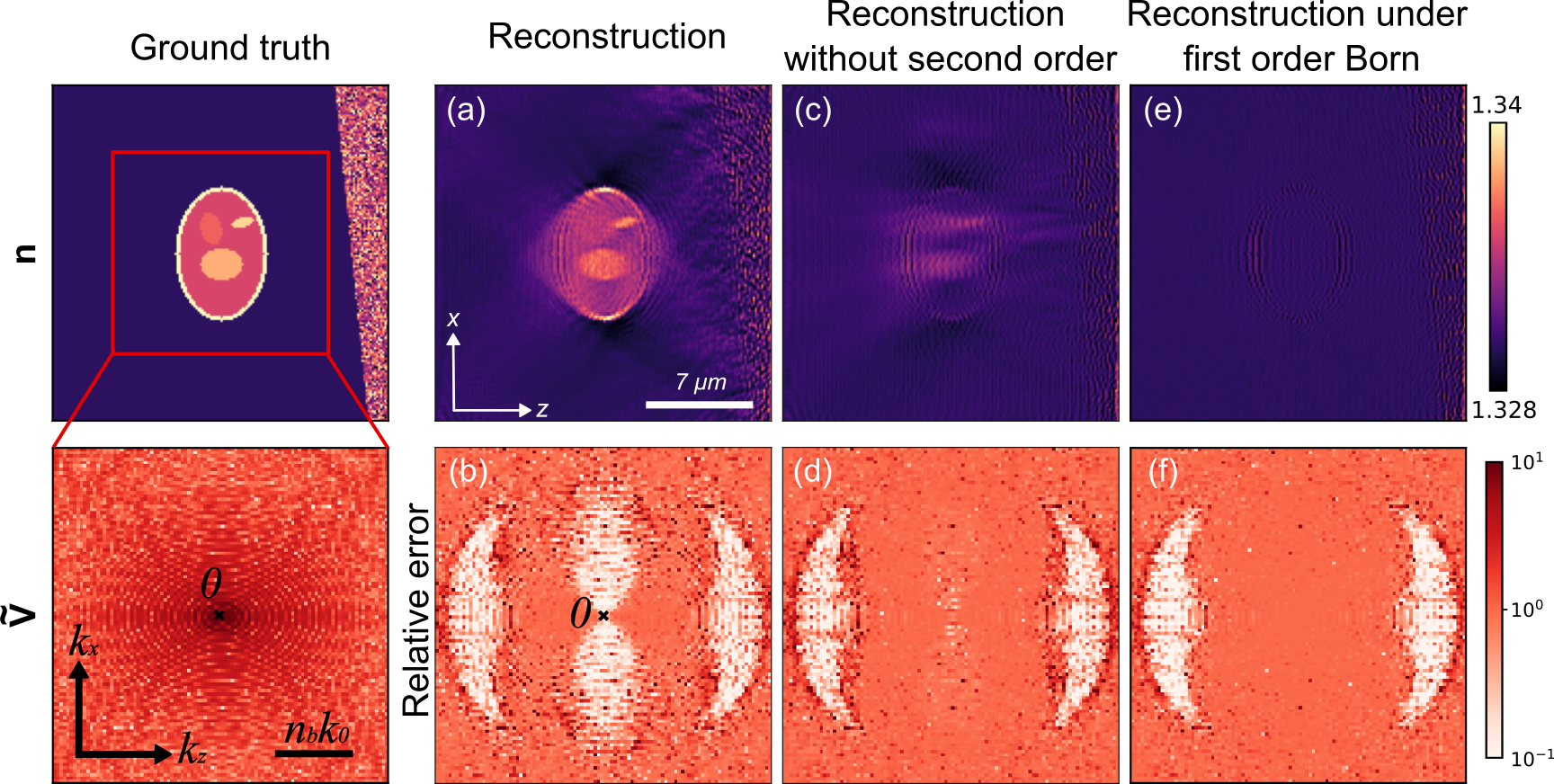}
    \caption{RI reconstruction and relative error of the reconstruction $\tilde{V}$ in Fourier space. The reconstruction are performed without regularization, with Adam optimizer and the temporal loss.
    (a,b) The ground truth is generated using the Lippmann–Schwinger model and the forward model used for reconstruction is the 5\textsuperscript{th}-order Born model.
    (c,d) The second-order (double scattering) term, including the cross terms $U_{12}$ and $U_{21}$, is removed from both the ground truth and the forward model.
    (e,f) Both the ground truth and the forward model are computed in the single-scattering regime using the first-order Born approximation.}
    \label{fig:fig10}
\end{figure}

As expected, using the 5\textsuperscript{th}-order Born model allows us to recover both the high and low-frequency components of the object in the foreground (Fig.~\ref{fig:fig10}(a,b)), which is consistent with the combination of multi-illumination transmission and reflection terms described in Coupland \textit{et al.}’s work \cite{coupland2008holography}.

When the second-order term is removed (Fig.~\ref{fig:fig10}(c,d)), the reconstruction quality decreases since the low-frequency information encoded in the cross terms disappears. Some of these components can still be partially recovered through higher-order terms, although their contribution remains limited, justifying the main contribution of second-order term in reconstruction.
Finally, removing all multiple-scattering terms (Fig.~\ref{fig:fig10}(e,f)) makes it impossible to retrieve the low-frequency information, even with a reflector in the background.

Moreover, Fig.~\ref{fig:fig10} provides an estimate of the maximum resolution, as the reconstruction can only recover information within a finite region of the object’s spatial Fourier space. Following Coupland \textit{et al.} \cite{coupland2008holography}, the lateral frequency range is bounded by $dk_x = |4 n_b k_0 \mathrm{NA}|$, while the axial range (corresponding to direct reflection information) extends over $dk_z = |4 n_b k_0|$.
For our microscopy configuration with incident wavelengths within $[800nm, 875nm]$, the maximum spatial resolution is about $150nm$. The figure shown here corresponds to a $140~nm$ resolution in order to clearly highlight the information which can be recovered in Fourier space.


\end{document}